\documentclass[a4paper]{article}
\usepackage{amsmath,amssymb,theorem,fullpage}
\usepackage[nosort]{cite}

\let\frac\undefined

\allowdisplaybreaks

\numberwithin{equation}{section}

\def\Maketitle{{\def\newpage{}\maketitle}}

\def\eq#1{\begin{equation}#1\end{equation}}
\long\def\subeq#1{\begin{subequations}#1\end{subequations}}

\def\Align#1{\begin{align}#1\end{align}}

\def\Aligned#1{\begin{aligned}#1\end{aligned}}
\def\Gather#1{\begin{gather}#1\end{gather}}
\def\Gathered#1{\begin{gathered}#1\end{gathered}}
\def\Multline#1{\begin{multline}#1\end{multline}}

\def\Cases#1{\begin{cases}#1\end{cases}}
\def\d{\partial}
\def\bd{\bar\partial}

\def\sign{\mathop{\rm sign}\nolimits}
\def\Res{\mathop{\rm Res}\limits}

\def\cA{{\cal A}}
\def\cF{{\cal F}}
\def\cDR{{\cal D}^R}
\def\cDRphys{{\cal D}^{R,\text{\rm phys}}}
\def\cDL{{\cal D}^L}
\def\cDLphys{{\cal D}^{L,\text{\rm phys}}}

\def\cT{{\cal T}}

\def\cW{{\cal W}}

\def\sh{\mathop{\rm sh}\nolimits}
\def\ch{\mathop{\rm ch}\nolimits}

\def\lcolon{\mathopen:}
\def\rcolon{\mathclose:}
\def\C{{\mathbb C}}
\def\R{{\mathbb R}}
\def\Z{{\mathbb Z}}

\def\llangle{\mathopen{\langle\!\langle}}
\def\rrangle{\mathclose{\rangle\!\rangle}}

\def\us{{\underline s}}
\def\e{{\rm e}}
\def\i{{\rm i}}

\def\tJ{{\tilde J}}

\def\bz{{\bar z}}

\def\bc{{\bar c}}
\def\bh{{\bar h}}
\def\bl{{\bar l}}
\def\bcA{{\bar{\cal A}}}
\def\lvac{\langle\text{\rm vac}|}
\def\rvac{|\text{\rm vac}\rangle}
\def\bfP{{\bf P}}
\def\bfQ{{\bf Q}}
\def\bfa{{\bf a}}

\def\fh{{\mathfrak h}}
\def\rad{{\text{\rm rad}}}
\def\sym{{\text{\rm sym}}}
\def\tLsym{{t_L^\sym}}

\makeatletter

\def\section{\@startsection{section}{1}{\z@}%
                                   {-3.5ex \@plus -1ex \@minus -.2ex}%
                                   {2.3ex \@plus.2ex}%
                                   {\normalfont\normalsize\bfseries}}
\def\subsection{\@startsection{subsection}{2}{\z@}%
                                     {-3.25ex\@plus -1ex \@minus -.2ex}%
                                     {1.5ex \@plus .2ex}%
                                     {\normalfont\normalsize\bfseries\itshape}}
\def\@seccntformat#1{\csname the#1\endcsname.~~}
\long\def\@makecaption#1#2{%
  \vskip\abovecaptionskip
  \sbox\@tempboxa{\small#1. #2}%
  \ifdim \wd\@tempboxa >0.9\hsize
  {\leftskip=0.05\hsize\rightskip=0.05\hsize\relax\small
    #1. #2\par}
  \else
    \global \@minipagefalse
    \hb@xt@\hsize{\hfil\box\@tempboxa\hfil}%
  \fi
  \vskip\belowcaptionskip}
\def\Appendix{\appendix
  \def\@seccntformat##1{Appendix~\csname the##1\endcsname.~~}}

\let\over\@@over
\let\atop\@@atop
\let\above\@@above
\let\overwithdelims\@@overwithdelims
\let\atopwithdelims\@@atopwithdelims
\let\abovewithdelims\@@abovewithdelims

\makeatother

\newtheorem{theorem}{Theorem}
\newtheorem{proposition}{Proposition}

\begin{document}

\title{\Large Form factors of descendant operators: $A^{(1)}_{L-1}$ affine Toda theory}
\author{Oleg Alekseev and Michael Lashkevich\\[\medskipamount]
{\normalsize\it\raggedright
Landau Institute for Theoretical Physics,
142432 Chernogolovka of Moscow Region, Russia}}
\date{}
\Maketitle

\begin{abstract}
In the framework of the free field representation we obtain exact form factors of local operators in the two-dimensional affine Toda theories of the $A^{(1)}_{L-1}$ series. The construction generalizes Lukyanov's well-known construction to the case of descendant operators. Besides, we propose a free field representation with a countable number of generators for the `stripped' form factors, which generalizes the recent proposal for the sine/sinh-Gordon model. As a check of the construction we compare numbers of the operators defined by these form factors in level subspaces of the chiral sectors with the corresponding numbers in the Lagrangian formalism. We argue that the construction provides a correct counting for operators with both chiralities. At last we study the properties of the operators with respect to the Weyl group. We show that for generic values of parameters there exist Weyl invariant analytic families of the bases in the level subspaces.
\end{abstract}

\section{Introduction}

The bootstrap approach to form factors in two-dimensional integrable quantum field theory makes it possible to calculate the form factors exactly by solving a set of difference equations for analytic functions called form factor axioms\cite{Karowski:1978vz,Smirnov:1984sx,Smirnovbook}. Any solution to these equations provides a local operator in the theory. The main conjecture of the approach is that vice versa the set of form factors of any local operator is a solution to the bootstrap equations. Though a very general approach to solving the form factor axioms was proposed by Smirnov\cite{Smirnovbook}, the problem of identification of the operators defined by the bootstrap form factors to the fields defined in the usual Lagrangian formalism is not solved in full generality. Moreover, the integral form of Smirnov's solution and many other proposals make it difficult to study them.

Here we consider the two-dimensional affine Toda models of the $A^{(1)}_{L-1}$ series\cite{Arinshtein:1979pb}. These models are models of an $(L-1)$-component real scalar field $\varphi(x)$ with an exponential interaction potential. In the particular case $L=2$, i.~e.\ of the sinh-Gordon model, different approaches for the form factors were developed~\cite{Smirnovbook,Koubek:1993ke,Lukyanov:1997bp,Babujian:1999ht,Babujian:2002fi,Feigin:2008hs}. For generic values of $L$ Babujian and Karowski~\cite{Babujian:2003za} proposed earlier a general form of the solution to the form factor axioms in this case in an integral form. We propose a solution in terms of finite sums based on Lukyanov's free field formalism for form factors~\cite{Lukyanov:1993pn}. Lukyanov~\cite{Lukyanov:1997yb} found the solutions to the bootstrap equations that correspond to the exponential operators $\e^{\i\alpha\varphi(x)}$ and completely identified them. Following the guidelines of~\cite{Feigin:2008hs} we find a representation for form factors of the so called descendant operators, i.~e. the operators of the form $(\d_{\mu_1}^{k_1}\varphi_{i_1})\ldots(\d_{\mu_r}^{k_r}\varphi_{i_r})\e^{\i\alpha\varphi}$. These operators may be considered as elements of the Fock spaces generated by modes of the field $\varphi(x)$ from the exponential operator in the radial quantization picture. Though up to now we are unable to identify these descendant operators with particular solutions to the bootstrap equations, we can identify some spaces of solutions with the Fock spaces over given exponential operators. Besides, in the case of the so called chiral descendants we are able to find bootstrap counterparts of the level subspaces of the Fock spaces.

An important feature of the affine Toda models is the existence of the so called reflection relations between the operators in the theory~\cite{Zamolodchikov:1995aa,Fateev:1997nn,Fateev:1997yg,Ahn:1999dz}. These relations connect operators with different values of $\alpha$ related by the action of the Weyl group. We prove the existence of these reflection relations for our solutions. Moreover, we show that there are analytic in $\alpha$ families of Weyl invariant bases in the Fock spaces. We hope this proof to be a step towards solution of the identification problem.

The paper is organized as follows. In Sec.~\ref{sec:Preliminaries} we describe the model, fix the notation and recall the main results of Lukyanov's free field representation. In Sec.~\ref{sec:CommutativeAlgebra} we introduce an auxiliary commutative algebra that allows us to generalize the free field representation to the descendant operators. We also cite several simple physical consequences of the construction and carry out the counting of the descendant operators defined by the bootstrap form factors. Sect.~\ref{sec:StrippedBosonization} is dedicated to an alternative free field construction, which is an important ingredient in the proof of the reflection relations in Secs.~\ref{sec:ReflectionExponential},~\ref{sec:ReflectionDescendant}. In Sec.~\ref{sec:ReflectionExponential} we use some recurrent relations to prove the explicit form of the reflection relations of the exponential operators, while in Sec.~\ref{sec:ReflectionDescendant} they are used to prove the existence of reflection relations for the descendant operators. The explicit reflection relations for the level~1 descendant operators are given in~Sec.~\ref{sec:LevelOne}.

\section{Preliminaries}
\label{sec:Preliminaries}

Let $\fh$ be the $(L-1)$-dimensional Cartan subalgebra of the simple Lie algebra $A_{L-1}$ and $\fh^*$ be its dual, the bracket $\langle\cdot,\cdot\rangle$ be either the Killing form restricted to $\fh$ or its dual on~$\fh^*$. Let $\alpha_i\in\fh^*$, $i=1,\ldots,L-1$ be the simple roots of the algebra, $\langle\alpha_i,\alpha_j\rangle=2\delta_{ij}-\delta_{i,j+1}-\delta_{i,j-1}$. Let $\alpha_0=-\sum^{L-1}_{i=1}\alpha_i$ be the affine root. Let $\rho$ be the half sum of the positive roots. In terms of the simple roots $\rho=\sum^{L-1}_{i=1}{i(L-i)\over2}\alpha_i$ and  $\langle\rho,\alpha_i\rangle=1$, $i>0$. Let $H_s$, $s=1,\ldots,L$ be the set of weights of the first fundamental representation $\pi_1$ of the algebra, $\langle\alpha_i,H_s\rangle=\delta_{is}-\delta_{i,s-1}$, so that $\alpha_i=H_i-H_{i+1}$.

Let $\varphi(x)\in\fh_\R$ be the real ($L-1$)-component field with the action
\eq{
S[\varphi]=\int d^2x\,\left({\langle\d_\mu\varphi,\d^\mu\varphi\rangle\over8\pi}
-{\mu\over2}\sum^{L-1}_{i=0}\e^{b\alpha_i\varphi}\right).
\label{TodaAction}
}
Let us stress that the sum in the r.~h.~s.\ contains the summation over all simple roots of the \emph{affine} Lie algebra $A^{(1)}_{L-1}$, including the affine root~$\alpha_0$. The theory is called the \emph{affine Toda field theory} associated with the algebra~$A^{(1)}_{L-1}$.

Below it will be convenient to use the letters $\omega$, $Q$, $p$ defined as follows:
\eq{
\omega=\e^{2\pi\i/L},
\qquad
Q=b+b^{-1},
\qquad
b=\sqrt{p\over1-p}.
\label{addnotation}
}
The parameter $p$ is always thought to be irrational.

We shall also use the light-cone variables and derivatives
$$
z=x^1-x^0,
\qquad
\bz=x^1+x^0,
\qquad
\d={\d\over\d z},
\qquad
\bd={\d\over\d\bz}.
$$
We adjusted the definition of the light-cone coordinates to fit the usual conformal field theory notation system.

The spectrum of the model consists of $L-1$ particles of masses\cite{Arinshtein:1979pb}
\eq{
M_k=[k]M_1,
\qquad
k=1,\ldots,L-1,
\label{particlemasses}
}
where we used the $q$-number type notation:
\eq{
[k]={\omega^{k/2}-\omega^{-k/2}\over\omega^{1/2}-\omega^{-1/2}}
={\sin{\pi k\over L}\over\sin{\pi\over L}}.
\label{[k]def}
}
The mass of the lightest particle $M_1$ is proportional to $\mu^{1/2(1-b^2)}$. The exact relation between $M_1$ and $\mu$ is known explicitly~\cite{Fateev:1993av}.

The space of local operators of the model consists of the exponential operators
\eq{
V_a(x)=\e^{Q(a+\rho)\varphi(x)}
\label{Vadef}
}
and their descendants, i.~e. linear combinations of the fields%
\footnote{We ignore the possible factor $\varphi^m$, since it can be obtained as the $m$-fold derivative in the variable~$a$.}%
\eq{
(\alpha_{i_1}\d^{l_1}\varphi)\cdots(\alpha_{i_r}\d^{l_r}\varphi)
(\alpha_{j_1}\bd^{\bl_1}\varphi)\cdots(\alpha_{j_s}\bd^{\bl_s}\varphi)\e^{Q(a+\rho)\varphi}(x)
\label{descendantsdef}
}
for any integers $r,s\ge0$ and $l_1,\ldots,\bl_s>0$. The pair of numbers $(l,\bl)$, defined as
\eq{
l=\sum^r_{p=1}l_p,
\qquad
\bl=\sum^s_{p=1}\bl_p,
\label{nnbardef}
}
is called the level of a descendant operator, while the numbers $l$ and $\bl$ separately will be referred to as chiral levels. The difference $S=l-\bl$ is the Lorentz spin of the operator, while the sum $D=Q^2\langle a+\rho,a+\rho\rangle+l+\bl$ is the scaling dimension in the ultraviolet region. The descendant operators with $\bl=0$ are called chiral, while those with $l=0$ are called antichiral.

In the radial quantization picture operators at some point, e.~g. $x=0$, are put in one-to-one correspondence to vectors of some auxiliary vector space. Namely, the field $\varphi(x)$ can be expanded in a kind of Laurent series in the vicinity of the point $x=0$:%
\footnote{Surely, this expansion only holds for very small vicinity of zero, $|z|,|z'|\ll M_1^{-1}$, where the field can be considered as a massless free boson field.}
\eq{
\alpha_i\varphi(x)=\bfQ_i-\i\bfP_i\log z\bz+\sum_{n\ne0}{\bfa_{in}\over\i n}z^{-n}
+\sum_{n\ne0}{\bar\bfa_{in}\over\i n}\bar z^{-n},
\qquad
i=1,\ldots,L-1,
\label{varphi-expansion}
}
where the operators $\bfQ_i$, $\bfP_i$, $\bfa_{in}$, $\bar\bfa_{in}$ form a Heisenberg algebra with the commutation relations
\eq{
[\bfP_i,\bfQ_j]=-\i\langle\alpha_i,\alpha_j\rangle,
\qquad
[\bfa_{im},\bfa_{jn}]=m\langle\alpha_i,\alpha_j\rangle\delta_{m+n,0},
\qquad
[\bar\bfa_{im},\bar\bfa_{jn}]=m\langle\alpha_i,\alpha_j\rangle\delta_{m+n,0}.
\label{PQaa-alg}
}
The operator $V_a(0)$ corresponds in this picture to the vector $|a\rangle_\rad$, such that
\eq{
\bfa_{in}|a\rangle_\rad=\bar\bfa_{in}|a\rangle_\rad=0\quad(n>0),
\qquad
\bfP_i|a\rangle_\rad=Q\langle\alpha_i,a+\rho\rangle|a\rangle_\rad.
\label{alpharad}
}
The radial vacuum vector $\rvac_\rad=|-\rho\rangle_\rad$ corresponds to the unit operator so that $|a\rangle_\rad=\e^{\i Q(a+\rho)\bfQ}\rvac_\rad$. Up to some $c$-number factors the operators (\ref{descendantsdef}) correspond to the vectors
\eq{
\bfa_{i_1,-l_1}\cdots\bfa_{i_r,-l_r}\bar\bfa_{j_1,-\bl_1}\cdots\bar\bfa_{j_s,-\bl_s}|a\rangle_\rad.
\label{descendants-vec}
}
These vectors span a Fock module with the highest weight vector $|a\rangle_\rad$, which can be written as a tensor product $\cF_a\otimes\bar\cF_a$ of two chiral components. The module $\cF_a$ is the Fock module spanned on the vectors (\ref{descendants-vec}) with $\bl=0$, while the module $\bar\cF_a$ is spanned on those with $l=0$. In turn, each of the modules can be split into a sum of the level subspaces. For example, $\cF_a\simeq\bigoplus^\infty_{l=0}\cF_{a,l}$, where $\cF_{a,l}$ is spanned by the level $l$ vectors. The dimensions of the subspaces $\cF_{a,l}$ are given by the well-known generating function:
\eq{
\sum^\infty_{l=0}q^l\dim\cF_{a,l}=\prod^\infty_{m=1}{1\over(1-q^m)^{L-1}}.
\label{cFngen}
}
Besides, there is a natural isomorphism $T:\cF_a\to\bar\cF_a$ due to the (time-reversal) map $\bfa_{in}\leftrightarrow\bar\bfa_{in}$. This isomorphism preserves the level: $T:\cF_{a,l}\to\bar\cF_{a,l}$.

For any vector $v\in\cF_a\otimes\bar\cF_a$ we define an operator $\Phi_a[v](x)$ as the operator corresponding to the state~$v|a\rangle_\rad$.
The reflection symmetry conjecture\cite{Zamolodchikov:1995aa,Fateev:1997nn,Fateev:1997yg,Ahn:1999dz} declares the following property. Let $\cW$ be the Weyl group of the $A_{L-1}$ simple Lie algebra. Then for a generic value of the parameter $a$ and any element $w\in\cW$ there exists a map $R_a(w):\cF\otimes\cF\to\cF\otimes\cF$, such that
\eq{
\Phi_{wa}[v](x)=\Phi_a[R_a(w)v](x).
\label{Reflection}
}
Since this property originates from the conformal field theory, it preserves the level of the operators:
$$
R_a(w)\cF_{al}\otimes\bar\cF_{a\bl}=\cF_{wa,l}\otimes\bar\cF_{wa,\bl}.
$$
Besides, it factorizes into chiral components:
\eq{
R_a(w)=r_a(w)\otimes Tr_a(w)T^{-1},
\qquad
r_a(w):\cF_a\to\cF_{wa}.
\label{Rfactorization-cft}
}

We shall see below, that there is another bijection between $\cF_a$ and $\bar\cF_{a'}$, which is natural from the point of view of the expressions for form factors. Let $w_*\in\cW$ be the element defined by the relation
\eq{
w_*\alpha_i=-\alpha_{L-i}
\quad\Leftrightarrow\quad
w_*H_s=H_{L+1-s}.
\label{w_*def}
}
The Weyl group is known to be generated by the reflections $w_i$ such that
$$
w_ia=a-\langle a,\alpha_i\rangle\alpha_i,
\qquad
i=1,\ldots,L-1.
$$
In terms of these generators the element $w_*$ is given by
$$
w_*=w_1(w_2w_1)(w_3w_2w_1)\cdots(w_{L-1}w_{L-2}\cdots w_1).
$$
Since $w_*^2=1$, we can define an automorphism of the Weyl group
\eq{
\widetilde w=w_*ww_*,
\label{w_*adj}
}
so that
$$
\widetilde{w_i}=w_{L-i}.
$$
For generic values of $a$ there exists a bijection $T^*_a:\cF_{w_*a}\to\bar\cF_a$, preserving the level, $T^*_a\cF_{w_*a,l}=\bar\cF_{a,l}$, such that the factorization property of the reflection operator reads
\eq{
R_a(w)=r_a(w)\otimes T^*_{wa}r_{w_*a}(\widetilde w)T_a^{*-1}.
\label{Rfactorization}
}
Such correspondence, seeming utterly artificial from the Lagrangian point of view, turns out to be a symmetry of the expressions for form factors.

Consider the exponential fields. Since $\dim\cF_0=1$, we have
\eq{
G_a^{-1}V_a(x)=G_{wa}^{-1}V_{wa}(x),
\qquad
\forall w\in\cW,
\label{exprefproperty}
}
where $G_a=\langle V_a(x)\rangle$ is the vacuum expectation value of the operator $V_a(x)$, which is known exactly\cite{Ahn:1999dz}.

Now we recall Lukyanov's representation~\cite{Lukyanov:1997yb} for form factors of exponential operators in the models~(\ref{TodaAction}). We do not need the detailed description in terms of the auxiliary free field, and we only formulate the result. Let us introduce the vertex operators $\Lambda_s(\theta)$, $s=1,\ldots,L$, with the two-point trace functions
\eq{
\Aligned{
\llangle\Lambda_s(\theta')\Lambda_s(\theta)\rrangle
&=R(\theta-\theta'),
\\
\llangle\Lambda_{s'}(\theta')\Lambda_s(\theta)\rrangle
&=R(\theta-\theta')F\left(\theta-\theta'+\sign(s-s'){\i\pi\over L}\right),
}\label{LambdaLambdaTr}
}
where
$$
\log R(\theta)
=-4\int{dt\over t}\,{\sh(L-1)t\,\sh pt\,\sh(1-p)t\over\sh^2Lt}\ch{L(\pi-\i\theta)t\over\pi}
$$
and
\eq{
F\left(\theta\pm{\i\pi\over L}\right)
={\sh\left({\theta\over2}\pm{\i\pi p\over L}\right)
\sh\left({\theta\over2}\pm{\i\pi(1-p)\over L}\right)
\over\sh{\theta\over2}\sh\left({\theta\over2}\pm{\i\pi\over L}\right)}.
\label{Ffuncdef}
}
Below we need generic multipoint trace functions and the normal product $\lcolon\cdots\rcolon$. We define both by the equations
\eq{
\Gathered{
\Lambda_{s_N}(\theta_N)\cdots\Lambda_{s_1}(\theta_1)
=\lcolon\Lambda_{s_N}(\theta_N)\cdots\Lambda_{s_1}(\theta_1)\rcolon
\prod_{1\le m<n\le N}\llangle\Lambda_{s_n}(\theta_n)\Lambda_{s_m}(\theta_m)\rrangle,
\\
\llangle\lcolon\Lambda_{s_N}(\theta_N)\cdots\Lambda_{s_1}(\theta_1)\rcolon\rrangle=1.
}\label{LambdaTraces}
}
Define the operators
\eq{
\Lambda_{s_1\ldots s_k}(\theta)
=\lcolon\prod^k_{j=1}\Lambda_{s_j}\left(\theta+{\i\pi(n+1-2j)\over L}\right)\rcolon,
\qquad
1\le s_1<\cdots<s_k\le L.
\label{Lambdaproddef}
}
In particular,
\eq{
\Lambda_{12\ldots L}(\theta)=1.
\label{Lambda12dotsL}
}
It is convenient to introduce the central element $\hat a$ with the values in $\fh^*$ and the trace function $\llangle\cdots\rrangle_a$ such that
$$
\llangle X(\hat a)\rrangle_a=\llangle X(a)\rrangle
$$
for any operator function $X(a)$. With this notation Lukyanov's generators are given by
\eq{
T_k(\theta)
=\lambda'_k\sum_{1\le s_1<\cdots<s_k\le L}
\omega^{\langle\hat a,H_{s_1\ldots s_k}\rangle}\Lambda_{s_1\ldots s_k}(\theta),
\qquad
H_{s_1\ldots s_k}=H_{s_1}+\cdots+H_{s_k}
\label{Tndef}
}
with the normalization constants
$$
\lambda'_k=\sqrt{L\over2\sin\pi p}
\exp\int{dt\over t}\,{\sh pt\,\sh(1-p)t\over\sh t\,\sh^2Lt}\Bigl(\sh^2kt+\sh^2(L-k)t\Bigr).
$$
The form factors of the exponential operators are given by\cite{Lukyanov:1997yb}
\eq{
\lvac V_a(0)|k_1\theta_1,\ldots,k_N\theta_N\rangle
\equiv G_af_a(\theta_1,\ldots,\theta_N)_{k_1\ldots k_N}
=G_a\llangle T_{k_N}(\theta_N)\ldots T_{k_1}(\theta_1)\rrangle_a.
\label{expff}
}

The functions $f_a$ are analytic functions of the parameters~$\theta_n$ with complicated analytic structure. Nevertheless, it is easy to see that they can be reduced to functions with simpler analytic properties. Indeed, it is easy to see that
\eq{
\Gathered{
f_a(\theta_1,\ldots,\theta_N)_{k_1\ldots k_N}
=J_{N,a}(\e^{\theta_1},\ldots,\e^{\theta_N})_{k_1,\ldots,k_N}
\prod^N_{n=1}\lambda'_{k_n}\prod^N_{m<n}R_{k_mk_n}(\theta_m-\theta_n),
\\
R_{kl}(\theta)=\prod^k_{i=1}\prod^l_{j=1}R\left(\theta+{\i\pi\over L}(k-l-2i+2j)\right).
}\label{Jdef}
}
The functions $J_{N,a}(x_1,\ldots,x_N)_{k_1\ldots k_N}$ are rational in the variables $x_n$ and symmetric with respect to permutations of the pairs $(k_n,x_n)$. We suppose that form factors of descendants operators possess the same structure with appropriate rational $J$ functions.

The functions $J_{N,a}$ possess the property
\eq{
J_{N,a}(x_1^{-1},\ldots,x_N^{-1})=J_{N,w_*a}(x_1,\ldots,x_N)
\label{invprop}
}
with the element $w_*$ defined in~(\ref{w_*def}). Indeed, if we substitute $x_n\to x_n^{-1}$ for all~$n$, the functions $F(\theta_m-\theta_n+\i\pi/L)$ and $F(\theta_m-\theta_n-\i\pi/L)$ trade their places. It is equivalent to the substitution $\Lambda_s(\theta)\to\Lambda_{L+1-s}(\theta)$. To adapt the factors $\omega^{\langle a,H_{s_1\ldots s_{k_i}}\rangle}$ to this situation we can use the identity $\langle a,H_s\rangle=\langle a,w_*H_{L+1-s}\rangle=\langle w_*a,H_{L+1-s}\rangle$. That is why the subscript $w_*a$ appears in the r.~h.~s.\ of~(\ref{invprop}). Surely, due to the reflection relation (\ref{exprefproperty}), which will be proven later, the element $w_*$ can be eliminated from the r.~h.~s., but the form (\ref{invprop}) will be convenient for generalization to descendant operators.

\textbf{Remark.} In the case $L=2$ this reduces to the sinh/sine-Gordon model. In this case we may set
$$
\alpha_1=-\alpha_0=\sqrt2,
\qquad
H_1=-H_2=\rho={1\over\sqrt2}.
$$
The relation to the notation of\cite{Feigin:2008hs} is given by substitutions:
$$
p\to-p,
\qquad
\omega^p\to\omega^{-1},
\qquad
b\to-{\i\beta\over\sqrt2},
\qquad
a\to\sqrt2a,
\qquad
\Lambda_1(\theta)\to\Lambda_-(\theta),
\qquad
\Lambda_2(\theta)\to\Lambda_+(\theta).
$$
The Weyl group contains the only nontrivial element $w_1$ such that $w_1a=-a$ and $w_*=w_1$.

\section{Descendant operators and a commutative algebra}
\label{sec:CommutativeAlgebra}

Define an algebra $\cA=\oplus^\infty_{l=0}\cA_l$ as the commutative algebra generated by elements $\langle\alpha_i,c_{-n}\rangle$ with positive integers~$n$, so that we shall operate with the symbols $c_{-n}$ as with vectors in~$\fh$. Each level $l$ subspace $\cA_l$ is spanned by the vectors $\prod c_{-n_i}$ such that $\sum n_i=l$. We shall also need another copy $\bcA$ of the algebra $\cA$ generated by the components of the vectors~$\bc_{-n}$. The canonical homomorphism from $\cA$ to $\bcA$ will be denoted by bar: $\overline{c_{-n}}=\bc_{-n}$.

Let us define a bracket on the algebra $\cA$:
\eq{
\left(\prod^\infty_{n=1}(u_nc_{-n})^{\mu_n},\prod^\infty_{n=1}(v_nc_{-n})^{\nu_n}\right)
=\prod^\infty_{n=1}\mu_n!\,\langle u_n,v_n\rangle^{\mu_n}\delta_{\mu_n\nu_n},
\qquad\forall u_n,v_n\in\fh^*.
\label{cAbracket}
}
It is not difficult to check the consistency of this definition.

Define the currents
\eq{
a_s(z)=\exp\left(H_s\sum^\infty_{n=1}c_{-n}\omega^{-{L+1-2s\over2}n}z^n\right).
\label{as(z)}
}
Let
\eq{
a_{s_1\ldots s_k}(z)=\prod^k_{j=1}a_{s_j}\left(\omega^{n+1-2s_j\over2}z\right),
\qquad
1\le s_1<\cdots<s_k\le L.
\label{aproddef}
}
From the fact that $\sum^L_{s=1}H_s=0$ it is easily deduced that
\eq{
a_{12\ldots L}(z)=1.
\label{aprod1}
}
Let
\eq{
b_{s_1\ldots s_k}(z)=a_{L+1-s_k,\ldots,L+1-s_1}(z).
\label{bdef}
}
Now we can define modified $T$-operators as
\eq{
\cT_k(\theta)
=\lambda'_k\sum_{1\le s_1<\cdots<s_k\le L}
\omega^{\langle\hat a,H_{s_1\ldots s_k}\rangle}\Lambda_{s_1\ldots s_k}(\theta)
a_{s_1\ldots s_k}(\e^\theta)\bar b_{s_1\ldots s_k}(\e^{-\theta}).
\label{cTndef}
}
For any $g\in\cA\otimes\bcA$ the functions
\eq{
f^g_a(\theta_1,\ldots,\theta_N)_{k_1\ldots k_N}
=\left(\llangle\cT_{k_N}(\theta_N)\ldots\cT_{k_1}(\theta_1)\rrangle_a,g\right)
\label{fgdef}
}
define an operator $V^g_a(z)$ by its form factors
\eq{
\lvac V^g_a(x)|k_1\theta_1,\ldots,k_N\theta_N\rangle
=G_af^g_a(\theta_1,\ldots,\theta_N)_{k_1\ldots k_N}.
\label{Vgdef}
}
The collection of all functions $f^g_a(\theta_1,\ldots,\theta_N)_{k_1\ldots k_N}$ with $N=0,1,2,\ldots$ will be denoted below as~$f^g_a$.

We also need the identity
\eq{
\left(a_{s_1}(x_1)\ldots a_{s_N}(x_N),\prod^\infty_{n=1}(u_nc_{-n})^{\mu_n}\right)
=\prod^\infty_{n=1}\left(\sum^N_{j=1}
\langle u_n,H_{s_j}\rangle\,\omega^{-{L+1-2s_j\over2}n}x_j^n\right)^{\mu_n}.
\label{aaprodbracket}
}
The rules (\ref{LambdaLambdaTr}) -- (\ref{LambdaTraces}) and (\ref{aaprodbracket}) are sufficient to explicitly calculate the form factors defined in~(\ref{fgdef}).

Now we are ready to extract the factor in the expression for form factors that is rational in the variables~$\e^{\theta_n}$, as we promised above. Indeed, one can write the form factors in the form
\eq{
\Gathered{
f^g_a(\theta_1,\ldots,\theta_N)_{k_1\ldots k_N}
=J^g_{N,a}(\e^{\theta_1},\ldots,\e^{\theta_N})_{k_1,\ldots,k_N}
\prod^N_{i=1}\lambda'_{k_i}\prod^N_{i<j}R_{k_ik_j}(\theta_i-\theta_j),
}\label{Jgdef}
}
where the functions $J^g_{N,a}(x_1,\ldots,x_N)_{k_1\ldots k_N}$ are rational functions of the variables~$x_1,\ldots,x_N$ symmetric with respect to permutations of the pairs $(k_i,x_i)$. Evidently, $J^1_{N,a}=J_{N,a}$. Though it is possible to write down these functions explicitly, their explicit expression is rather cumbersome and not too useful. In the next section we describe a free field representation that allows one to study these functions effectively.

Consider any element $g\in\cA_n\otimes\cA_{\bar n}$. The spin of the corresponding local operator $V^g_a$ is evidently equal to $n-\bar n$ due to the form factor axioms~\cite{Smirnovbook}. The situation with the (ultraviolet) conformal dimension is more complicated. From some experience in scaling limits of the lattice models we may expect that, if we normalize the form factors in a physical way, the element $g$ brings the factor~$M_1^{n+\bar n}$. It means that the element $g$ increases the scaling dimension by $n+\bar n$ in comparison with that of the exponential operator. But we cannot control mixing in any operators of lesser dimensions and the same spin. We may conclude that the elements $g\in\cA_n$ correspond to chiral level $n$ (spin $n$) descendants, while the elements $g\in\bcA_{\bar n}$ correspond to antichiral level $\bar n$ (spin $-\bar n$) descendants. As far as generic $g$ are concerned, we may only say that they correspond to a linear combination of the level $(l,\bar l)$ descendants with $l\le n$, $\bar l\le\bar n$. Vice versa, any level $(l,\bar l)$ descendant operator is a linear combination of the operators $V^g_a$ with $g$ beings elements of the subspaces $\cF_n\otimes\bar\cF_{\bar n}$ with $n\le l$, $\bar n\le\bar l$. This conclusion will be supported by the properties described below in Subsec.~\ref{subsec:IMs}--\ref{subsec:counting}.

Note that the definition (\ref{bdef}) has been chosen in such a way that the resulting functions $J^g_{N,a}$ satisfy the relation very similar to (\ref{invprop}) for $J_{N,a}$:
\eq{
J^{h\bh'}_{N,a}(x_1^{-1},\ldots,x_N^{-1})=J^{h'\bh}_{N,w_*a}(x_1,\ldots,x_N)
\label{invgprop}
}
for $h,h'\in\cA$. Though this property is not very important by itself, it will be helpful for obtaining formulas in the anti-chiral sector immediately from those for the chiral sector.

The form factors $f_a^g$ also possess two periodicity properties. Let $A=\sum^{L-1}_{i=1}i\alpha_i$. Then
\Align{
f_{a+A}^g(\theta_1,\ldots,\theta_N)_{k_1\ldots k_N}
&=\omega^{k_1+\cdots+k_N}f_a^g(\theta_1,\ldots,\theta_N)_{k_1\ldots k_N},
\label{fAperiodicity}
\\
f_{a+L\alpha_0}^g(\theta_1,\ldots,\theta_N)_{k_1\ldots k_N}
&=f_a^g(\theta_1,\ldots,\theta_N)_{k_1\ldots k_N}.
\label{falpha0periodicity}
}
The sense of these properties can be clarified in the case $g=1$. Consider the Weyl group transformations $w_\pm$:
$$
w_+=w_-^{-1}=w_1w_2\cdots w_{L-1}.
$$
These transformations act on the affine roots $\alpha_i$, $i=0,1,\ldots,L-1$ as cyclic permutations: $w_\pm\alpha_i=\alpha_{i\pm1}$. The cyclic permutations are symmetry transformations of the action~(\ref{TodaAction}):
$$
S[w_+\varphi]=S[\varphi].
$$
Therefore, under the action of $w_+$ the form factors must be invariant:
$$
w_+(\langle0|V_a(0)|k_1\theta_1\ldots k_N\theta_N\rangle)
=\langle0|V_a(0)|k_1\theta_1\ldots k_N\theta_N\rangle.
$$
The breather states are transformed as follows%
\footnote{In fact, the overall sign in the exponent of $\omega$ can be assumed arbitrarily. Change of this sign only redefines $k_i\to L-k_i$.}
$$
w_\pm(|k_1\theta_1\ldots k_N\theta_N\rangle)
=\omega^{\pm(k_1+\cdots+k_N)}|k_1\theta_1\ldots k_N\theta_N\rangle.
$$
This can be extracted, for example, from the quasiclassic picture of the eigentones in the vicinity of the minimum of the potential~$U(\varphi)={\mu\over2}\sum^{L-1}_{i=0}\e^{b\alpha_i\varphi}$.

It is easy to check that
$$
w_+\rho=\rho+A+L\alpha_0,
\qquad
w_-\rho=\rho-A.
$$
Hence,
$$
w_+(V_a(x))=\e^{Q(a+\rho)(w_+\varphi(x))}=\e^{Q(w_-(a+\rho))\varphi(x)}
=\e^{Q(w_-a-A+\rho)\varphi(x)}=V_{w_-a-A}(x).
$$
We have
$$
\omega^{k_1+\cdots+k_N}\langle0|V_{w_-a-A}(0)|k_1\theta_1\ldots k_N\theta_N\rangle
=\langle0|V_a(0)|k_1\theta_1\ldots k_N\theta_N\rangle.
$$
With the reflection relation (\ref{exprefproperty}) for the exponential operators it is consistent with (\ref{fAperiodicity}) and~(\ref{falpha0periodicity}). Thus, the periodicity properties reflect, in a sense, the cyclic symmetry of the $A^{(1)}_{L-1}$ Dynkin diagram.

Now let us cite several important properties of the expressions for form factors obtained above.

\subsection{Integrals of motion}
\label{subsec:IMs}

Let
\eq{
\iota_n=\sum^{L-1}_{i=1}\omega^{{L-i\over2}n}[in](\alpha_ic_{-n}),
\qquad
\iota_{-n}=\overline{\iota_n},
\qquad n\in\Z_{>0}\setminus L\Z_{>0}.
\label{hIdef}
}
It is straightforward to check that, according to (\ref{aaprodbracket}), the element $\iota_n$ produces a common factor in all terms in (\ref{fgdef}) related to the sums~(\ref{cTndef}), resulting in the identity
\eq{
f_a^{\iota_ng}(\theta_1,\ldots,\theta_N)_{k_1\ldots k_N}
=\left(\sum^N_{m=1}{[k_mn]\over[n]}\e^{n\theta_m}\right)
f_a^g(\theta_1,\ldots,\theta_N)_{k_1\ldots k_N},
\qquad
n\in\Z\setminus L\Z.
\label{fhIg}
}
for any $g\in\cA\otimes\bcA$. In the factor in parentheses, one can recognize the eigenvalue of the (appropriately normalized) spin $n$ integral of motion~$I_n$. It means that
\eq{
V^{\iota_ng}_a(x)=[V^g_a(x),I_n].
\label{VIcommut}
}
This result is consistent with~\cite{Niedermaier:1992pb,Niedermaier:1993sq}. Note also that the set of the integrals of motion $I_n$ is the natural deformation of the set of the integrals of motion in the massless (conformal) limit, and the commutators (\ref{VIcommut}) add the value $n$ to the spin of the operator and the value $|n|$ to the chiral (antichiral) level of the descendant operator for $n>0$ ($n<0$). It means that if we could identify some operator $V^g_a$ with a descendant operator in the Lagrangian formulation, we would have a large subspace of $\cF_a\otimes\bar\cF_a$ that consists of the operators generated from $V^g_a$ by the integrals of motion identified.

\subsection{Factorization property}
\label{subsec:factorization}

Let $h,h'\in\cA$. Consider the $\Lambda\to\infty$ asymptotics of the function
$$
f^{h\bh'}_a(\theta_1,\ldots,\theta_M,\theta_{M+1}+\Lambda,\ldots,\theta_N+\Lambda)
_{k_1\ldots k_Mk_{M+1}\ldots k_N}.
$$
It is easy to check that $F(\theta\pm\Lambda),R(\theta\pm\Lambda)\to1$ in this limit. Let $\us_i=(s_i^{(1)},\ldots,s_i^{(k_i)})$ be sets of integers $1\le s_i^{(1)}<\cdots<s_i^{(k_i)}\le L$. From (\ref{aaprodbracket}) we immediately get
$$
\Gathered{
(a_{\us_1}(x_1)\ldots a_{\us_M}(x_M)a_{\us_{M+1}}(x_{M+1}\e^\Lambda)\ldots a_{\us_N}(x_N\e^\Lambda),h)
=(a_{\us_{M+1}}(x_{M+1}\e^\Lambda)\ldots a_{\us_N}(x_N\e^\Lambda),h),
\\
(a_{\us_1}(x_1)\ldots a_{\us_M}(x_M)
a_{\us_{M+1}}(x_{M+1}\e^{-\Lambda})\ldots a_{\us_N}(x_N\e^{-\Lambda}),h)
=(a_{\us_1}(x_1)\ldots a_{\us_M}(x_M),h)
}
$$
for $h\in\cA$. Finally, we obtain the following \textit{asymptotic factorization property}\cite{Delfino:2005wi}:
\Multline{
f^{h\bh'}_a(\theta_1,\ldots,\theta_M,\theta_{M+1}+\Lambda,\ldots,\theta_N+\Lambda)
_{k_1\ldots k_Mk_{M+1}\ldots k_N}
\\
=f^h_a(\theta_{M+1}+\Lambda,\ldots,\theta_N+\Lambda)_{k_{M+1}\ldots k_N}
f^{\bh'}_a(\theta_1,\ldots,\theta_M)_{k_1\ldots k_M}
\text{ as $\Lambda\to+\infty$}.
\label{factorization}
}
This property means that it is always possible to extract chiral parts of descendants using this limit. Roughly speaking, the high velocity right moving particles only know about the chiral part of a local operator, while the high velocity left moving particles only know about its antichiral part.

From the factorization property together with (\ref{invgprop}) we immediately obtain that if the reflection property holds, it should possess the factorized form~(\ref{Rfactorization}). In other words, if it is possible define an action of $R_a(w)$ on the algebra $\cA\otimes\bcA$ such that $V^{R_a(w)h\bh'}_a(x)=V^{h\bh'}_a(x)$, we necessarily have
\eq{
V^{h\bh'}_a(x)=V^{(r_a(w)h)(\overline{r_{w_*a}(\widetilde w)h'})}_a(x),
\label{Vrefl}
}
where $r_a(w)$ is the restriction of $R_a(w)$ on $\cA\simeq\cA\otimes\bcA_0$. The possibility to define such action will be proven in Sec.~\ref{sec:ReflectionDescendant}.

Another consequence concerns the identification of operators. For any $h\in\cA_n$ and $h'\in\cA_{\bar n}$ define the operator $\mathcal V^{h\bh'}_a(x)=M_1^{n+\bar n}V^{h\bh'}_a(x)$. The operators $\mathcal V^h_a$ and $\mathcal V^{\bh'}_a$ are level $(n,0)$ and level $(0,\bar n)$ descendants correspondingly. Moreover, they must be linear combinations of the vectors of the form (\ref{descendantsdef}) with $\mu$-independent coefficients. Hence, their form factors are proportional to $G_aM_1^n\propto M_1^{Q^2\langle\rho+a,\rho+a\rangle+n}$ and $G_aM_1^{\bar n}\propto M_1^{Q^2\langle\rho+a,\rho+a\rangle+\bar n}$. It follows from (\ref{factorization}) and (\ref{Vgdef}) that the leading term in the operator $\mathcal V_a^{h\bh'}$ is proportional to $G_aM_1^{n+\bar n}\propto M_1^{Q^2\langle\rho+a,\rho+a\rangle+n+\bar n}$. It means that the operator $V_a^{h\bh'}$ is a nonzero level $(n,\bar n)$ descendant plus some operators of lesser dimensions.

\subsection{Descendants counting}
\label{subsec:counting}

We want to prove that, for generic values of $a$, the operators $V^g_a$ with different values of $g$ differ. For this purpose we first prove this fact for a particular asymptotics in~$a$. Then we apply the deformation argument.

Let
$$
a(\tau)={L\tau\over2\pi\i}\rho,
\qquad
a(\tau)H_s={L\tau\over2\pi\i}\left({L+1\over2}-s\right).
$$
We shall consider the limit $\tau\to+\infty$. Evidently,
$$
{2\pi\i\over L\tau}a(\tau)H_{12\ldots k}={k(N-k)\over2},
\qquad
{2\pi\i\over L\tau}a(\tau)H_{s_1s_2\ldots s_k}<{k(N-k)\over2}
\quad\text{for $s_k>k$.}
$$
Therefore
\eq{
\e^{-\tau{k(L-k)\over2}}T_k(z)|_{\hat a=a(\tau)}
=\Lambda_{12\ldots k}(z)+O(\e^{-\tau})\quad\text{as $\tau\to+\infty$.}
\label{Tktauasymp}
}
Then the functions $J^g_{N,a}$ have the following asymptotics:
\eq{
\left.\e^{-{\tau\over2}\sum^N_{i=1}k_i(L-k_i)}
J^g_{N,a}(x_1,\ldots,x_N)_{k_1\ldots k_N}\right|_{\tau\to\infty}
=P^g(X_1|\ldots|X_{L-1}),
\qquad
X_k=\{x_i|k_i=k\}.
\label{Jgtauasymp}
}
The functions $P^g$ are polynomials defined by the following relations:
\subeq{\label{Pdef}
\Gather{
P^{g_1g_2}=P^{g_1}P^{g_2},
\qquad
P^{C_1g_1+C_2g_2}=C_1P^{g_1}+C_2P^{g_2}
\qquad(\forall g_1,g_2\in\cA,\ C_1,C_2\in\C).
\label{Palg}
\\
P^{\alpha_ic_{-n}}(X_1|\ldots|X_{L-1})
=\omega^{i-1\over2}S_n(X_i)+\sum^{L-1}_{k=i+1}\omega^{i-k\over2}(1-\omega)S_n(X_k),
\label{Pcdef}
\\
P^{\alpha_{L-i}\bc_{-n}}(X_1|\ldots|X_{L-1})
=-\omega^{-{i-1\over2}}S_{-n}(X_i)+\sum^{L-1}_{k=i+1}\omega^{k-i-1\over2}(1-\omega)S_{-n}(X_k),
\label{Pbcdef}
}}
Here
$$
S_n(x_1,\ldots,x_N)=\sum^N_{i=1}x_i^n.
$$
The products of polynomials $S_n$ for $n>0$ of a given power form a basis of the symmetric polynomials of the respective power for a large enough number of variables.

First, consider the case $g\in\cA$. Since we are interested in the whole collections of form factors rather than the form factors for particular numbers of particles, we may consider the functions $z_{in}=S_n(X_i)$, $i=1,\ldots,L-1$ as independent variables. The equation (\ref{Pcdef}) defines a map from the algebra $\cA$ to the algebra of polynomials in the variables~$z_{in}$. This map is invertible. Indeed, Eq.~(\ref{Pcdef}) makes it possible to express any monomial $z_{in}$ in terms of $P^{\alpha_ic_{-n}}$ and the monomials $z_{jn}$, $j>i$. Applying it recursively we can express any monomial $z_{in}$ in terms of a linear combination of the polynomials $P^{\alpha_jc_{-n}}$ with $j\ge i$. Hence, $z_{in}=P^{g_{in}}$, where $g_{in}=\sum^{L-1}_{j=i}A_j\alpha_jc_{-n}$ with some uniquely defined coefficients~$A_j$. This defines a map from the algebra of polynomials in the variables $z_{in}$ to the algebra~$\cA$.

This proves that different elements $g_1\ne g_2\in\cA$ produce different polynomials $P^{g_1}\ne P^{g_2}$ of the variables~$z_{kn}$. Since the form factors are analytic functions of the variable $a$, these elements produce different collections of form factors~$f^{g_1}_a\ne f^{g_2}_a$.

Now suppose that $g=\sum h_i\bh'_i$, where $\{h_i\},\{h'_i\}\subset\cA_l$ are sets of linearly independent elements. Suppose that $f^g_a=0$. Then due to the factorization property the combination
$$
\sum f^{h_i}_a(\theta_1,\ldots,\theta_M)f^{\bh'_i}_a(\theta'_1,\ldots,\theta'_N)=0.
$$
This contradicts to the linear independence of the form factors $f^{h_i}$ for generic~$a$. This proves the

\begin{theorem}
For generic values of~$a$ the linear map $g\mapsto f^g_a$ from $\cA\otimes\bcA$ into the space of collections of functions is an injection, i.~e.\ it is invertible as a map onto its image.
\end{theorem}

As an immediate consequence we have the

\begin{proposition}\label{descounting}
For generic values of~$a$ the dimension of the space of the operators $V^g_a$ with $g\in\cA_l\otimes\bcA_\bl$ is equal to the dimension of the corresponding subspace of the Fock space $\dim(\cF_l\otimes\cF_\bl)=\dim\cF_l\cdot\dim\cF_\bl$. The dimensions of the spaces of the operators, $V^g_a$ with $g\in\cA_l$ or $g\in\bcA_l$ are equal to that of the subspace, $\dim\cF_l$.
\end{proposition}

For chiral (antichiral) operators Proposition~\ref{descounting} means that the conjecture that the chiral (antichiral) descendants are the operators $V_a^g(x)$ with $g\in\cA$ ($g\in\bcA$) is consistent with the operator counting from the bosonic picture (\ref{descendantsdef}),~(\ref{nnbardef}).

\section{The stripped bosonization}
\label{sec:StrippedBosonization}

To prove the reflection property for the descendant operators we shall need a free field representation of the functions~$J^g_{N,a}$, which are obtained from the functions $f^g_a$ by stripping out the $R$ factors. We shall call it stripped bosonization. This bosonization differs from that described in\cite{Lukyanov:1993pn,Lukyanov:1997yb} in that, first, the Heisenberg algebra is generated by a countable set of elements rather than a continuous one and, second, the functions $J^g_{N,a}$ for all $g\in\cA\otimes\bcA$ are expressed in terms of matrix elements rather than traces. The price paid for these advantages is that the residue of the kinematic pole is not a $c$-number, but a new vertex operator. We shall see below that this new vertex operator will be an important ingredient of our proof.

Consider the Heisenberg algebra with the generators $d^{(s)}_n$, $s=1,\ldots,L$, $n\in\Z$, $n\ne0$ and the commutation relations
\eq{
[d^{(s)}_m,d^{(s)}_n]=0,
\qquad
[d^{(s')}_m,d^{(s)}_n]=m\delta_{m+n,0}A^{\sign(s'-s)}_n
\quad(s'\ne s)
\label{ddcommut}
}
with
\eq{
A^\pm_n=(\omega^{\mp pn}-\omega^{\mp n})(1-\omega^{\pm pn}).
\label{Apmdef}
}
Note that
\eq{
A^-_n=A^+_{-n}=\omega^nA^+_n.
\label{Apmrel}
}
Let $\hat a$ be an additional central element. Define the vacuums $|1\rangle_a$ and ${}_a\langle1|$ by the relations
\eq{
d^{(s)}_n|1\rangle_a=0,
\quad
\hat a|1\rangle_a=a|1\rangle_a,
\quad
{}_a\langle1|d^{(s)}_{-n}=0,
\quad
{}_a\langle1|\hat a={}_a\langle1|a,
\quad
{}_a\langle1|1\rangle_a=1\quad(n>0)
\label{dvacuum}
}
and let $\lcolon\cdots\rcolon$ be the corresponding normal ordering operation. We shall also write
$$
\langle\ldots\rangle_a\equiv{}_a\langle1|\ldots|1\rangle_a.
$$
The Fock space generated by the operators $d^{(s)}_n$, $n>0$, from the vacuum ${}_a\langle1|$ will be denoted as~$\cDR_a$, while that generated by $d^{(s)}_{-n}$, $n>0$, from the vacuum $|1\rangle_a$ will be denoted as~$\cDL_a$. They admit a natural grading $\cDR_a=\bigoplus^\infty_{n=0}\cDR_{a,n}$, $\cDL_a=\bigoplus^\infty_{n=0}\cDL_{a,n}$ so that $\cDR_{a,m}d^{(s)}_n\subseteq\cDR_{a,m+n}$, $d^{(s)}_n\cDL_{a,m}\subseteq\cDL_{a,m-n}$.

Let us introduce the vertex operators
\eq{
\lambda_s(z)=\exp\sum_{n\ne0}{d^{(s)}_n\over n}z^{-n}.
\label{lambdasdef}
}
Note that the exponential in the r.~h.~s.\ does not need any normal ordering due to commutativity of all elements $d^{(s)}_n$ with a given~$s$. It is easy to check that
\eq{
\Aligned{
\lambda_s(z')\lambda_s(z)
&=\lcolon\lambda_s(z')\lambda_s(z)\rcolon,
\\
\lambda_{s'}(z')\lambda_s(z)
&=\lambda_s(z)\lambda_{s'}(z')
=f\Bigl({z\over z'}\Bigr)\lcolon\lambda_{s'}(z')\lambda_s(z)\rcolon,
\qquad
s'>s,\quad{z\over z'}\ne 1,\omega.
}\label{lambdalambdaprods}
}
Here
\eq{
f(z)=F\left(\log z-{\i\pi\over L}\right)={(z-\omega^p)(z-\omega^{1-p})\over(z-1)(z-\omega)},
\qquad
f(z)=f(\omega/z).
\label{f(z)def}
}
Let
\eq{
\lambda_{s_1\ldots s_k}(z)
=\lcolon\prod^k_{m=1}\lambda_{s_m}\left(z\omega^{k+1-2m\over2}\right)\rcolon,
\qquad
1\le s_1<\cdots<s_k\le L.
\label{lambdaproddef}
}
Note that the vertex operator $\lambda_{12\ldots L}(z)$ is not equal to one and plays an important role below. Its important property is
\eq{
\lambda_{12\ldots L}(z)\lambda_s(x)
=\prod^{L-1}_{m=1}f\left({z\over x}\omega^{L+1-2m\over2}\right)
\lcolon\lambda_{12\ldots L}(z)\lambda_s(x)\rcolon.
\label{lambda12...Llambdaprod}
}
It is necessary to stress that the coefficient in the r.~h.~s.\ is $s$ independent.

Now define the stripped $W$ algebra currents
\eq{
t_k(z)=\sum_{1\le s_1<\cdots<s_k\le L}\omega^{\langle a,H_{s_1\ldots s_k}\rangle}
\lambda_{s_1\ldots s_k}(z).
\label{tndef}
}
It is evident that
\eq{
J_{N,a}(x_1,\ldots,x_N)_{k_1\ldots k_N}
=\langle t_{k_N}(x_N)\ldots t_{k_1}(x_1)\rangle_a\>.
\label{Jt...t}
}
To obtain the functions $J^g_{N,a}$ for an arbitrary element $g$ we need an additional construction. Consider two representations of the algebra $\cA$ in the Heisenberg algebra, $\pi_R$ and $\pi_L$, defined as follows:
\eq{
\Aligned{
\pi_R(\alpha_ic_{-n})
&=R^{(i)}_n={\omega^{-{L+1-2i\over2}n}\over A^+_n}(d^{(i)}_n-d^{(i+1)}_n),
\\
\pi_L(\alpha_ic_{-n})
&=L^{(i)}_n={\omega^{-{L+1-2i\over2}n}\over A^+_n}(d^{(L-i)}_{-n}-d^{(L+1-i)}_{-n}).
}\label{piRLdef}
}
These operators satisfy the commutation relations
\subeq{\label{picommut}
\Gather{
[\pi_R(\alpha_ic_{-n}),\lambda_s(z)]
=\langle\alpha_i,H_s\rangle\omega^{-{L+1-2s\over2}n}z^n\lambda_s(z),
\label{piRcommut}
\\
[\lambda_s(z),\pi_L(\alpha_ic_{-n})]
=\langle\alpha_i,H_{L+1-s}\rangle\omega^{{L+1-2s\over2}n}z^{-n}\lambda_s(z),
\label{piLcommut}
\\
[\pi_R(\alpha_ic_{-m}),\pi_L(\alpha_jc_{-n})]
=m\delta_{mn}(A^+_m)^{-1}(\delta_{i+j,L-1}+\omega^m\delta_{i+j,L+1}-(1+\omega^m)\delta_{i+j,L}).
\label{piRLcommut}
}}
Define the `physical' vectors
\eq{
{}_a\langle h|={}_a\langle1|\pi_R(h),
\qquad
|h\rangle_a=\pi_L(h)|1\rangle_a.
\label{hstatedef}
}
We call them `physical' due to the following reason. Consider the functions
\eq{
\tJ^{h\bh'}_{N,a}(x_1,\ldots,x_N)_{k_1\ldots k_N}
={}_a\langle h|t_{k_1}(x_1)\ldots t_{k_N}(x_N)|h'\rangle_a.
\label{Jtildedef}
}
These functions define the form factors, which will be denoted as~$\tilde f^{h\bh'}_a$ according to the same equation as~(\ref{Jgdef}) and can be expressed in terms of the $J$ functions. Indeed, we can push the element $\pi_R(h)$ from the definition of the vector ${}_a\langle h|$ to the right and the element $\pi_L(h')$ from the vector $|h\rangle_a$ to the left. Due to the commutation relation (\ref{piRcommut}) and (\ref{piLcommut}) we get just the functions from the r.~h.~s.\ of~(\ref{aaprodbracket}), which enter the definition of~$J^{h\bh'}_a$. The only complication is the appearance of some extra terms due to the commutation relation~(\ref{piRLcommut}).

Precisely, the relation between the functions $\tJ$ and $J$ is as follows. Introduce two maps
\eq{
\pi_{LR}(h\bh')=\pi_L(h')\pi_R(h),
\qquad
\pi_{RL}(h\bh')=\pi_R(h)\pi_L(h').
\label{piRLLR}
}
These maps are bijections of $\cA\otimes\bcA$ to the subalgebra of the universal enveloping of the Heisenberg algebra generated by the elements $\pi_L(c_{-n})$ and~$\pi_R(c_{-n})$, $n>0$. Then
\eq{
\tJ^{h\bh'}_{N,a}(x_1,\ldots,x_N)_{k_1,\ldots,k_N}
=J^{\pi_{LR}^{-1}\circ\pi_{RL}(h\bh')}_{N,a}(x_1,\ldots,x_N)_{k_1,\ldots,k_N}.
\label{JtildeJrel}
}
More explicitly, take the product $\pi_R(h)\pi_L(h')$ and push these two factors through each other. We get a combination of the form
$$
\pi_R(h)\pi_L(h')=\sum_i\pi_L(h'_i)\pi_R(h_i).
$$
Then
$$
\tJ^{h\bh'}_{N,a}(x_1,\ldots,x_N)_{k_1,\ldots,k_N}
=\sum_i J^{h_i\bh'_i}_{N,a}(x_1,\ldots,x_N)_{k_1,\ldots,k_N}.
$$
The most important feature of this expression is that the function $\tJ^{h\bh'}$ with $h\in\cA_l$, $h'\in\cA_\bl$ is expressed in terms of the functions $J^{h_i\bh'_i}$ with $h_i\in\cA_{l_i}$, $h'_i\in\cA_{\bl_i}$ such that $l_i\le l$, $\bl_i\le\bl$ and vice versa. It means, in particular, that the factorization property (\ref{factorization}) holds as well for the form factors corresponding to the $\tilde J$ functions.

These `physical' vectors form the `physical' subspaces in the spaces $\cDR_a$ and~$\cDL_a$:
\eq{
\Aligned{
\cDRphys_{a,n}
&=\Bigl\{{}_a\langle h|\>\Big|\>h\in\cA_n\Bigr\},
\qquad
\cDRphys_a=\bigoplus^\infty_{n=0}\cDRphys_{a,n},
\\
\cDLphys_{a,n}
&=\Bigl\{|h\rangle_a\>\Big|\>h\in\cA_n\Bigr\},
\qquad
\cDLphys_a=\bigoplus^\infty_{n=0}\cDLphys_{a,n}.
}\label{Dphysdef}
}
Evidently,
\eq{
\dim\cDRphys_{a,n}=\dim\cDLphys_{a,n}=\dim\cF_n.
\label{physdim}
}

It is convenient to find a definition of the subspaces $\cDRphys$, $\cDLphys$ as kernels of some operators. Introduce a set of operators $D_n$ ($n\ne0$) such that
\eq{
[D_n,\pi_R(h)]=[D_n,\pi_L(h)]=0.
\label{Dncommut}
}
They are given by
\eq{
D_n=\sum^L_{s=1}\omega^{-{L+1-2s\over2}n}d^{(s)}_n.
\label{Dndef}
}
These elements commute with each other:
\eq{
[D_m,D_n]=0.
\label{DDcommut}
}
Note that
\eq{
\lambda_{1\ldots L}(z)=\exp\sum_{n\ne0}{D_nz^{-n}\over n}.
\label{lambda1...L-D}
}
Let ${}_a\langle U|$, $|V\rangle_a$ be some states from the Fock modules over ${}_a\langle1|$, $|1\rangle_a$. It can be shown that
\eq{
\Aligned{
{}_a\langle U|D_{-n}=0\quad\forall n>0
&\qquad\Leftrightarrow\qquad
\exists h\in\cA:\quad
{}_a\langle U|={}_a\langle1|\pi_R(h),
\\
D_n|V\rangle_a=0\quad\forall n>0
&\qquad\Leftrightarrow\qquad
\exists h\in\cA:\quad
|V\rangle_a=\pi_L(h)|1\rangle_a.
}\label{Dnproperty}
}
Therefore, the `physical' subspaces can be also defined as
\eq{
\Aligned{
\cDRphys_{a,n}
&=\Bigl\{{}_a\langle v|\in\cDR_a\>\Big|\>{}_a\langle v|D_{-m}=0\ \forall m>0\Bigr\},
\\
\cDLphys_{a,n}
&=\Bigl\{|v\rangle_a\in\cDL_a\>\Big|\>D_m|v\rangle_a=0\ \forall m>0\Bigr\},
}\label{physdef}
}

\section{Recurrent relations and reflection property for exponential operators}
\label{sec:ReflectionExponential}

Our first step in the proof of the reflection property is to prove it for the form factors of the exponential operators~$V_a(x)$. Since the expressions (\ref{expff}), (\ref{Jt...t}) are not explicitly invariant under the transformations of the Weyl group, we need another representation for the $J_{N,a}$ functions. It turns out that such a Weyl invariant representation can be found in the form of a recurrent relation in the number of particles $N$ starting form the explicitly invariant expression for~$N=0$. In this section we set $I=(1,\ldots,N)$, $X=(x_1,\ldots,x_N)$. Besides, we use the notation $\hat I_n=I\setminus\{n\}$, $\hat X_n=X\setminus\{x_n\}$.

It follows from Eq.~(\ref{Restkt1}) below that any function $J_{N,a}(\ldots)_{k_1\ldots k_N}$ can be expressed in terms of the function $J_{\sum k_i,a}(\ldots)_{1\ldots1}$. Hence, it is sufficient to obtain a recurrent relation for any subset of the $J$ functions that contains the functions with $k_1=\cdots=k_N=1$. We choose the subset that consists of the functions with an arbitrary $k_1$ and with fixed $k_2=\cdots=k_N=1$. Namely, consider the function
\eq{
J_{k,N+1,a}(z;X)
=\prod^N_{n=1}\prod_{m=1}^{k-1}f^{-1}\left({z\over x_i}\omega^{k+1-2m\over2}\right)
\langle t_k(z)t_1(x_1)\ldots t_1(x_N)\rangle_a,
\label{JkN+1(zX)}
}
which will be considered as an analytic function of the variable $z$, while the other variables~$X$ will be considered as parameters.

The prefactor in this expression cancels redundant poles. Indeed, consider the product of two vertex operators $t_k(z)t_1(x)$. It possesses four simple (see Appendix~\ref{app:simplepoles}) poles at the points
\eq{
z=x\omega^{\pm{k+1\over2}},x\omega^{\pm{k-1\over2}},
\label{tkt1SimplePoles}
}
$(k-2)$ double poles at the points
\eq{
z=x\omega^{k+1-2m\over2},\quad m=2,\ldots,k-1,
\label{tkt1Poles}
}
and $2(k-1)$ simple zeros at the points
\eq{
z=x\omega^{\pm\left(p-{k+1-2m\over2}\right)}\quad m=1,\ldots,k-1.
\label{tkt1Zeros}
}
Two more simple zeros have no fixed position and depend on a matrix element. The product
$$
\prod_{m=1}^{k-1}f^{-1}\left({z\over x}\omega^{k+1-2m\over2}\right)
=\prod^{k-1}_{m=1}{(z-x\omega^{k+1-2m\over2})(z-x\omega^{-{k+1-2m\over2}})
\over(z-x\omega^{p-{k+1-2m\over2}})(z-x\omega^{-p+{k+1-2m\over2}})}
$$
just cancels all fixed zeros and all poles except those at $z=x\omega^{\pm(k+1)/2}$. Thus, the only poles of the product
$$
\prod_{m=1}^{k-1}f^{-1}\left({z\over x}\omega^{k+1-2m\over2}\right)t_k(z)t_1(x)
$$
are located at $z=x\omega^{\pm(k+1)/2}$, where the residues are proportional to $t_{k+1}(x)$:
\eq{
\Res_{z=x\omega^{\pm{k+1\over2}}}
\prod_{m=1}^{k-1}f^{-1}\left({z\over x}\omega^{k+1-2m\over2}\right)t_k(z)t_1(x)
=\pm x\omega^{\pm{k+1\over2}}\kappa_p\,
t_{k+1}(x\omega^{\pm{k\over2}}),
\label{Restkt1}
}
where
\eq{
\kappa_p={(\omega^{1-p}-1)(\omega^p-1)\over\omega-1}
={2\i\sin{\pi p\over L}\sin{\pi(1-p)\over L}\over\sin{\pi\over L}}.
\label{kappadef}
}
It means that the prefactor in (\ref{JkN+1(zX)}) essentially reduces the number of poles of the resulting function.

The case $k=L-1$ is a special one since both poles coincide: $x\omega^{\pm L/2}=-x$. From the physical point of view, it corresponds to a kinematical rather than a dynamical pole. The residue is given by
\eq{
\Res_{z=-x}
\prod_{m=1}^{L-2}f^{-1}\left(-{z\over x}\omega^{-m}\right)t_{L-1}(z)t_1(x)
=x\kappa_p\left(t_L(-x\omega^{1/2})-t_L(-x\omega^{-1/2})\right).
\label{Restkt1kin}
}
It is important that $t_L(z)$ cannot be extracted from the operator products of $t_k$ for $k<L$. In spite of its apparent triviality, it is a new substance and this fact will be essentially used in the next section.

The prefactor in (\ref{JkN+1(zX)}) makes the analytic structure of the function $J_{k,N+1,a}(z;X)$ to be simple. Its only dynamic poles are located at $z=x_i\omega^{\pm(k+1)/2}$. The equation (\ref{Restkt1}) makes it possible to calculate residues at these poles:
\eq{
\Res_{z=x_n\omega^{\pm{k+1\over2}}}J_{k,N+1,a}(z;X)
=\pm x_n\omega^{\pm{k+1\over2}}R^\pm_{N,n}(X)J_{k+1,N,a}(x_n\omega^{\pm{k\over2}};\hat X_n),
\label{ResJ}
}
where
\eq{
R^\pm_{N,n}(X)
=\kappa_p
\prod_{m\in\hat I_n}f\left(\left(x_m\over x_n\right)^{\pm1}\right).
\label{RNdef}
}
For the special case $k=L-1$ we have
\eq{
\Res_{z=-x_n}J_{L-1,N+1,a}(z;X)
=x_n\left(R^-_{N,n}(X)-R^+_{N,n}(X)\right)J_{1,N-1,a}(\hat X_n),
\label{ResJL-1}
}
since $J_{L,N,a}(x;\hat X_n)=J_{1,N-1,a}(\hat X_n)$ due too~(\ref{lambda12...Llambdaprod}). Fortunately, there is no need to consider this case separately while deriving the recurrent relations. It will be taken into account implicitly by the cyclic property~(\ref{Jcyclic}) below.

One can separate the pole contributions from the regular part:
\Align{
J_{k,N+1,a}(z;X)
&=J^{(\infty)}_{k,N+1,a}(z;X)
\notag
\\*
&\quad+\sum^N_{n=1}{x_n\omega^{k+1\over2}\over z-x_n\omega^{k+1\over2}}
\>R^+_{N,n}(X)J_{k+1,N,a}(x_n\omega^{k\over2};\hat X_n)
\notag
\\*
&\quad-\sum^N_{n=1}{x_n\omega^{-{k+1\over2}}\over z-x_n\omega^{-{k+1\over2}}}
\>R^-_{N,n}(X)J_{k+1,N,a}(x_n\omega^{-{k\over2}};\hat X_n),
\label{Jinfexp}
}
where the function $J^{(\infty)}_{k,N+1,a}(z;X)$ is regular in the variable $z$ everywhere except the points $z=0$ and $z=\infty$. Note, that the sum over the poles is of the order $O(z^{-1})$ as $z\to\infty$. Thus, the asymptotic behavior of the function $J_{k,N+1,a}(z;X)$ as a function of $z$ is governed by~$J^{(\infty)}_{k,N+1,a}(z;X)$:
\eq{
J_{k,N+1,a}(z;X)-J^{(\infty)}_{k,N+1,a}(z;X)=O(z^{-1})\quad\text{as $z\to\infty$}.
\label{JJinfdif}
}
We may use another expansion
\Align{
J_{k,N+1,a}(z;X)
&=J^{(0)}_{k,N+1,a}(z;X)
\notag
\\*
&\quad-\sum^N_{n=1}{x^{-1}_n\omega^{-{k+1\over2}}\over z^{-1}-x_n^{-1}\omega^{-{k+1\over2}}}
\>R^+_{N,n}(X)J_{k+1,N,a}(x_n\omega^{k\over2};\hat X_n)
\notag
\\*
&\quad+\sum^N_{n=1}{x^{-1}_n\omega^{{k+1\over2}}\over z^{-1}-x_n^{-1}\omega^{{k+1\over2}}}
\>R^-_{N,n}(X)J_{k+1,N,a}(x_n\omega^{-{k\over2}};\hat X_n),
\label{J0exp}
}
where the function $J^{(0)}_{k,N+1,a}(z;X)$ is again regular everywhere except $z=0,\infty$. It is evident that the behavior of the function $J_{k,N+1,a}(z,X)$ in the vicinity of the point $z=0$ is governed by $J^{(0)}_{k,N+1,a}(z;X)$:
\eq{
J_{k,N+1,a}(z;X)-J^{(0)}_{k,N+1,a}(z;X)=O(z)\quad\text{as $z\to0$}.
\label{JJ0dif}
}
We shall use the notation
\eq{
D_{k,N,a}(X)
=\sum^N_{n=1}R^+_{N,n}(X)J_{k+1,N,a}(x_n\omega^{k\over2};\hat X_n)
-\sum^N_{n=1}R^-_{N,n}(X)J_{k+1,N,a}(x_n\omega^{-{k\over2}};\hat X_n).
\label{DkNa}
}
From the definitions (\ref{Jinfexp}), (\ref{J0exp}) it is easy to derive that
\eq{
J^{(0)}_{k,N+1,a}(z;X)-J^{(\infty)}_{k,N+1,a}(z;X)=D_{N,a}(X).
\label{J0infdif}
}
This relation shows that the both functions are nearly the same except the zero mode in~$z$. It means that it is sufficient establish the singular parts of $J^{(0)}$ and $J^{(\infty)}$ in the vicinity of $z=0$ and $z=\infty$ correspondingly and the zero mode of one of them.

What have been said above pertains equally to form factors of exponential and descendant operators. Now we want to restrict ourselves to the exponential operators. In order to fix the functions $J^{(\infty)}_{k,N+1,a}(z;X)$ and $J^{(0)}_{k,N+1,a}(z;X)$ we have to calculate the asymptotics of the function $J_{k,N+1,a}(z;X)$ as $z\to0$ and $z\to\infty$. Since $f(0)=f(\infty)=1$, it is finite in these limits and we have
\eq{
J^{(0)}_{k,N+1,a}(z;X)=J^{(\infty)}_{k,N+1,a}(z;X)=K_{k,a}J_{1,N,a}(X),
}
where
\eq{
K_{k,a}=J_{k,1,a}(0)=J_{k,1,a}(\infty)
=\sum_{1\le s_1<\cdots<s_n\le L}\omega^{\langle a,H_{s_1\ldots s_n}\rangle}.
\label{Kadef}
}
The fact that $D_{k,N,a}=0$ for the exponential operators provides a nontrivial identity
\eq{
\sum^N_{n=1}R^+_{N,n}(X)J_{k+1,N,a}(x_n\omega^{k\over2};\hat X_n)
=\sum^N_{n=1}R^-_{N,n}(X)J_{k+1,N,a}(x_n\omega^{-{k\over2}};\hat X_n).
\label{RRidentity}
}
We arrive to the
\begin{theorem}
\label{RecurrentRelation}
The recurrent relations
\Align{
J_{k,N+1,a}(z;X)
&=K_{k,a}J_{1,N,a}(X)
\notag
\\*
&\quad+\sum^N_{n=1}{x_n\omega^{k+1\over2}\over z-x_n\omega^{k+1\over2}}
\>R^+_{N,n}(X)J_{k+1,N,a}(x_n\omega^{k\over2};\hat X_n)
\notag
\\*
&\quad-\sum^N_{n=1}{x_n\omega^{-{k+1\over2}}\over z-x_n\omega^{-{k+1\over2}}}
\>R^-_{N,n}(X)J_{k+1,N,a}(x_n\omega^{-{k\over2}};\hat X_n)
\label{Jrec}
}
together with the initial conditions
\eq{
J_{k,0,a}=1,\qquad J_{k,1,a}(z)=K_{k,a}
\label{Jinit}
}
and with the cyclic property
\eq{
J_{L,N+1,a}(z;X)=J_{1,N,a}(X)
\label{Jcyclic}
}
uniquely define the set of functions~$J_{k,N,a}$.

Here the functions $R^\pm_{N,n}(X)$ are defined in~(\ref{RNdef}) and the factor $K_{k,a}$ is defined in~(\ref{Kadef}).
\end{theorem}

Due to (\ref{Restkt1}) any function $J_{N,a}(X)_{k_1,\ldots,k_N}$ with arbitrary $k_i$ can be constructed from certain function (\ref{JkN+1(zX)}) by taking residues at the dynamic poles. Hence, the recurrent relations (\ref{Jrec})--(\ref{Jcyclic}) provide a general construction for the form factors of exponential operators.

Note, that the function $K_{k,a}$ is nothing but the character of the fundamental representation $\pi_k$:
\eq{
K_{k,a}={\rm tr}_{\pi_k}\omega^{\langle a,H\rangle}.
\label{Ktrace}
}
Below we make use of the explicit formula
\eq{
K_{k,\lambda\alpha_i-\rho}
=4\omega^{ik}[k]\sin{\pi\lambda\over L}\sin{\pi(\lambda-1)\over L}.
\label{Kspec}
}
Now let us turn our attention to the reflection property for exponential fields. The reflection property (\ref{exprefproperty}) is an immediate consequence of theorem
\begin{theorem}
The functions $J^a_{k,N}$ are symmetric symmetric under the transformations of the Weyl group ${\cal W}$ of the Lie algebra $A_{L-1}${\rm:}
\eq{
J_{k,N,a}(z;X)=J_{k,N,wa}(z,X)
\qquad
\forall w\in\cW
\label{JkNareflection}
}
or, equivalently,
\eq{
J_{N,a}(x_1,\ldots,x_N)_{k_1,\ldots,k_N}
=J_{N,wa}(x_1,\ldots,x_N)_{k_1,\ldots,k_N}
\qquad
\forall w\in\cW.
\label{Jreflection}
}
\end{theorem}

Let us prove the theorem. The parameter $a$ only enters the recurrent relations (\ref{Jrec}), (\ref{Jinit}) in terms of~$K_{k,a}$. Hence, it is sufficient to prove the reflection property for this function. Since the Weyl group is generated by the simple reflections $w_i$, it is sufficient to prove the reflection property with respect to $w_i$, that is
$$
K_{k,a}=K_{k,w_ia}.
$$
It is easy to check from the definitions that
$$
\langle w_ia,H_s\rangle
=\Cases{\langle a,H_{s+1}\rangle,&\text{if $s=i$},\\
\langle a,H_{s-1}\rangle,&\text{if $s=i+1$},\\
\langle a,H_s\rangle&\text{otherwise}.}
$$
Therefore
$$
\Aligned{
\langle w_ia,H_{s_1\ldots s_m\ldots s_k}\rangle
&=\langle a,H_{s_1\ldots,s_m+1,\ldots s_k}\rangle,
&&\text{if $s_m=i$, $s_{m+1}>i+1$,}
\\
\langle w_ia,H_{s_1\ldots s_m\ldots s_k}\rangle
&=\langle a,H_{s_1\ldots,s_m-1,\ldots s_k}\rangle,
&&\text{if $s_m=i+1$, $s_{m-1}<i$,}
\\
\langle w_ia,H_{s_1\ldots s_k}\rangle
&=\langle a,H_{s_1\ldots s_k}\rangle
&&\text{otherwise}.
}
$$
Thus, any simple reflection acts as a permutation of the set $\{\langle a,H_{s_1\ldots s_k}\rangle\}$ of functions of the variable~$a$. Since the sum in (\ref{Kadef}) runs over the whole set, the function $K_{k,a}$ is invariant under simple Weyl reflections.

As an example of an application of the recurrent relations (\ref{Jrec})---(\ref{Jcyclic}) we prove the equation of motion for quantum fields in Appendix~\ref{app:EquationOfMotion}.

\section{Reflection property for descendant operators}
\label{sec:ReflectionDescendant}

The proof of the reflection property repeats in the main features that for the sine/sinh-Gordon model\cite{Feigin:2008hs}. The idea of the proof stems from the conjecture of\cite{Fateev:2006js}, developed further in\cite{Fateev:2009kp}, that all form factors can be obtained from those of the primary operators as coefficients of large $\theta$ expansions.

\begin{theorem}
For generic values of $a$ there exists a representation of the Weyl group $r_a$ on the algebra $\cA$ such that for any $h,h'\in\cA$ the equation holds
\eq{
\tilde J^{h\bh'}_{N,a}(x_1,\ldots,x_N)_{k_1\ldots k_N}
=\tilde J^{(r_a(w)h)(\overline{r_{w_*a}(\widetilde w)h'})}_{N,wa}(x_1,\ldots,x_N)_{k_1\ldots k_N}.
\label{Jhh'reflection}
}
\end{theorem}

The first step of the proof is to prove that the whole Fock space of the Heisenberg algebra (\ref{ddcommut}) can be spanned on vectors created by products of the operators~$t_k(x)$, $1\le k\le L$. More precisely, consider the expansion
\eq{
{}_a\langle1|t_{k_1}(\xi_1^{-1}z)\cdots t_{k_K}(\xi_K^{-1}z)
=\sum^\infty_{n=0}z^{-n}{}_a\langle n;k_1\xi_1,\ldots,k_K\xi_K|.
\label{ntstates}
}
For shortness, we shall write $\Xi=(k_1,\xi_1,\ldots,k_K,\xi_K)$. We want to prove that for generic values of~$a$ and large enough values of $K$ one can choose a set $\Xi^{(i)}$, $i=1,\ldots,\dim\cDR_n$, such that the vectors ${}_a\langle n;\Xi^{(i)}|$ form a basis in the space~$\cDR_n$. First, let us prove it in the limit $a=a(\tau)$, $\tau\to+\infty$ already used in Subsection~\ref{subsec:counting}. In this limit
\eq{
\e^{-\tau{k(L-k)\over2}}t_k(z)|_{\hat a=a(\tau)}=\lambda_{12\ldots k}(z)+O(\e^{-\tau})
\quad\text{as $\tau\to\infty$,}
\label{tktauasymp}
}
which is the full analog of~(\ref{Tktauasymp}). Therefore
\Multline{
\e^{-{\tau\over2}\sum^K_{i=1}k_i(L-k_i)}
{}_{a(\tau)}\langle1|t_{k_1}(\xi_1^{-1}z)\cdots t_{k_K}(\xi_K^{-1}z)
={}_{a(\tau)}\langle1|\lambda_{1\ldots k_1}(\xi_1^{-1}z)\cdots\lambda_{1\ldots k_K}(\xi_K^{-1}z)
+O(\e^{-\tau})
\\
=F(\xi_2/\xi_1,\ldots,\xi_K/\xi_1){}_{a(\tau)}\langle1|
\lcolon\lambda_{1\ldots k_1}(\xi_1^{-1}z)\cdots\lambda_{1\ldots k_K}(\xi_K^{-1}z)\rcolon
+O(\e^{-\tau}),
\label{stateasymp}
}
where $F(z_2,\ldots,z_K)$ is a product of functions $f$ of appropriate arguments. The particular form of this product is not essential for our purposes. Let us calculate the state in the last line:
\eq{
{}_{a(\tau)}\langle1|
\lcolon\lambda_{1\ldots k_1}(\xi_1^{-1}z)\cdots\lambda_{1\ldots k_K}(\xi_K^{-1}z)\rcolon
={}_{a(\tau)}\langle1|\exp\sum^\infty_{n=1}\sum^L_{s=1}{\kappa^{(s)}_nd^{(s)}_nz^{-n}\over n},
\label{statenormal}
}
where
\eq{
\kappa^{(s)}_n=\sum^K_{\scriptstyle i=1\atop\scriptstyle k_i\ge s}\omega^{{k_i+1-2s\over2}n}\xi_i^n.
\label{kappa(s)ndef}
}
Consider the expansion
$$
{}_{a(\tau)}\langle1|
\lcolon\lambda_{1\ldots k_1}(\xi_1^{-1}z)\cdots\lambda_{1\ldots k_K}(\xi_K^{-1}z)\rcolon
=\sum^\infty_{n=0}z^{-n}\>{}_{(-)}\langle n;\Xi|.
$$
Then
\eq{
{}_{(-)}\langle n;\Xi|
={}_{a(\tau)}\langle1|\sum^n_{r=1}\sum_{n_1,\ldots,n_r>0\atop n_1+\cdots+n_r=n}
C_{n_1\ldots n_r}\prod^r_{j=1}\sum^L_{s=1}\kappa_{n_j}^{(s)}d_{n_j}^{(s)}
\label{nXiform}
}
with some positive constants $C_{n_1\ldots n_r}$. It means that all possible products of $d^{(s)}_n$ enter the r.~h.~s.

For large enough cardinal numbers $\#\{i|k_i=s\}$, $1\le s\le L$, the functions $\kappa^{(s)}_{n'}$, $1\le s\le L$, $1\le n'\le n$ are functionally independent and can be considered as independent variables. Besides, the monomials $\kappa^{(s_1)}_{n_1}\cdots\kappa^{(s_r)}_{n_r}$ are linearly independent. Hence, for any nonzero set of numbers $A^{s_1\ldots s_r}_{n_1\ldots n_r}$, $r=1,\ldots,n$, $n_1,\ldots,n_r>0$, $n_1+\cdots+n_r=n$, we have
$$
\sum_r\sum_{s_1,\ldots,s_r\atop n_1,\ldots,n_r}\overline{A^{s_1\ldots s_r}_{n_1\ldots n_r}}
\kappa^{(s_1)}_{n_1}\cdots\kappa^{(s_r)}_{n_r}\ne0
$$
for some values of the variables~$\kappa^{(s)}_{n'}$. Therefore, the vector generated by the numbers $A^{s_1\ldots s_r}_{n_1\ldots n_r}$ is not orthogonal to at least one vector generated by the numbers $\kappa^{(s_1)}_{n_1}\cdots\kappa^{(s_r)}_{n_r}$. It means that there is no vector in~$\cDR_n$ orthogonal to all of the vectors generated by the products of $\kappa^{(s_j)}_{n_j}$. It proves that there exists a basis ${}_{(-)}\langle l;\Xi^{(I)}|$ in the space~$\cDR_l$. Now the deformation argument proves the same statement for generic values of~$a$.

Now we begin the second step of the proof. Let ${}_a\langle l;I|={}_a\langle l;\Xi_l^{(I)}|$ be a basis in the space~$\cDR_l$ related to any particular set of values of the parameters $\{\Xi^{(I)}_l|I=1,\ldots,\dim\cDR_l\}$. From the reflection property for the exponential operators (\ref{Jreflection}) we immediately conclude that for any nonnegative integers $l,\bl$ we have
$$
{}_a\langle l;I|t_{k_1}(x_1)\ldots t_{k_N}(x_N)|\bl;J\rangle_a
={}_{wa}\langle l;I|t_{k_1}(x_1)\ldots t_{k_N}(x_N)|\bl;J\rangle_{wa}
\qquad\forall w\in\cW.
$$
This identity provides a map $r_a(w):\cDR_l\to\cDR_l$ such that $r_a(w)({}_a\langle l;I|)={}_{wa}\langle l;I|$. Note that the left subscript $a$ at the vectors is essential, since the element of $\cDR_l$ generated by $\Xi^{(I)}_l$ depends on its value. Now our aim is to prove that this map is consistent with the restriction~(\ref{Dnproperty}), which selects `physical' states generated by~(\ref{piRLdef}).

Let the vectors ${}_a\langle1|\pi_R(h_{a,l,\mu})={}_a\langle\widetilde{l;\mu}|=\sum_Iv^\mu_I(a)\>{}_a\langle l;I|$ form a basis in the subspace $\cDRphys_{a,l}$. Similarly, let $\pi_L(h'_{a,\bl,\nu})|1\rangle_a=|\widetilde{\bl,\nu}\rangle_a=\sum_J\bar v^\nu_J|\bl,J\rangle_a$. Due to (\ref{lambda1...L-D}) we have
\Multline{
{}_a\langle1|t_{k_1}(x_1)\ldots t_{k_M}(x_M)D_nt_{k_{M+1}}(x_{M+1})\ldots t_{k_N}(x_N)|1\rangle_a
\\
={}_{wa}\langle1|t_{k_1}(x_1)\ldots t_{k_M}(x_M)
D_nt_{k_{M+1}}(x_{M+1})\ldots t_{k_N}(x_N)|1\rangle_{wa}.
\notag
}
We have
$$
0={}_a\langle\widetilde{l;\mu}|D_{-n}|l-n;J\rangle_a
=\sum_Iv^\mu_I(a)\>{}_a\langle l;I|D_{-n}|l-n;J\rangle_a
=\sum_Iv^\mu_I(a)\>{}_{wa}\langle l;I|D_{-n}|l-n;J\rangle_{wa}.
$$
Therefore,
$$
\sum_Iv^\mu_I(a)\>{}_{wa}\langle l;I|D_{-n}=0
$$
and there exists an element $h^w_{wa,l,\mu}$ such that
$$
{}_{wa}\langle1|\pi_R(h^w_{wa,l,\mu})=\sum_Iv^\mu_I(a)\>{}_{wa}\langle l;I|.
$$
Similarly, there exists an element $h^{\prime w}_{wa,\bl,\nu}$ such that
$$
\pi_L(h^{\prime w}_{wa,\bl,\nu})|1\rangle_{wa}
=\sum_J\bar v^\nu_J(a)|\bl;J\rangle_{wa}.
$$
Finally, we find
\Multline{
\langle\pi_R(h_{a,l,\mu})t(x_1)\ldots t(x_N)\pi_L(h'_{a,\bl,\nu})\rangle_a
={}_a\langle\widetilde{l;\mu}|t_{k_1}(x_1)\ldots t_{k_N}(x_N)|\widetilde{\bl,\nu}\rangle_a
\\
=\sum_{I,J}v^\mu_I(a)\bar v^\nu_J(a)\>
{}_a\langle l;I|t_{k_1}(x_1)\ldots t_{k_N}(x_N)|\bl,J\rangle_a
=\sum_{I,J}v^\mu_I(a)\bar v^\nu_J(a)\>
{}_{wa}\langle l;I|t_{k_1}(x_1)\ldots t_{k_N}(x_N)|\bl,J\rangle_{wa}
\\
=\langle\pi_R(h^w_{wa,l,\mu})t_{k_1}(x_1)\ldots t_{k_N}(x_N)
\pi_L(h^{\prime w}_{wa,\bl,\nu})\rangle_{wa}.
\notag
}
Thus, we obtained the map $r_a(w)$ on the subspaces $\cDRphys_a$,~$\cDLphys$:
$$
r_a(w)({}_a\langle1|\pi_R(h_{a,l,\mu}))={}_{wa}\langle1|\pi_R(h^w_{wa,l,\mu}),
\qquad
r_a(w)(\pi_L(h'_{a,\bl,\nu})|1\rangle_a)=\pi_L(h^{\prime w}_{wa,\bl,\nu})|1\rangle_{wa}.
$$
Comparing with the property~(\ref{invgprop}), which holds for the $\tilde J$ functions as well as for the~$J$ ones, we can define
$$
r_a(w)h_{a,l,\mu}=h^w_{wa,l,\mu},
\qquad
r_{w_*a}(\widetilde w)h'_{a,\bl,\nu}=h^{\prime w}_{wa,\bl,\nu}.
$$
Again this proves the factorized form~(\ref{Rfactorization}) of the reflection map.

{\bf An alternative construction.} It is easy to derive the commutation relation
\eq{
[D_n,t_k(z)]
=A_n{[kn]\over[n]}
z^nt_k(z),
\label{Dntk-commut}
}
where the coefficients
$$
A_n=-A^+_n\sum^L_{s=2}\omega^{-{L+1-2s\over2}n}
=\Cases{(-)^n\omega^{n/2}A^+_n,&n\not\in L\Z,\\
(1-L)A^+_n,&n\in L\Z.}
$$
are all nonzero for irrational values of~$p$. Besides, the ratio
$$
{[kn]\over[n]}=\sum^k_{s=1}\omega^{{k+1-2s\over2}n}
$$
is well defined (and equal to $k$) for $n\in L\Z$.

The commutation relation (\ref{Dntk-commut}) means that the product (\ref{ntstates}) satisfy the relations
\eq{
{}_a\langle1|t_{k_1}(\xi_1^{-1}z)\ldots t_{k_K}(\xi_K^{-1}z)D_{-n}=0,
\qquad
1\le n\le l,
\label{t...tD=0}
}
subject to the equations
\eq{
\sum^K_{m=1}{[k_mn]\over[n]}\,\xi_m^n=0,
\qquad
1\le n\le l,
\label{xieqs}
}
are satisfied. Hence,
\eq{
{}_a\langle n;k_1\xi_1,\ldots,k_K\xi_K|D_{-n'}=0,
\qquad
1\le n\le l,
\qquad
n'\ge1.
\label{ntphys}
}
If we also define
\eq{
t_{\bar k_1}(\eta_1z)\ldots t_{\bar k_{\bar K}}(\eta_Kz)|1\rangle_a
=\sum^\infty_{n=1}z^n|n;\bar k_1\eta_1,\ldots,\bar k_{\bar K}\eta_{\bar K}\rangle_a,
\label{ntstatebar}
}
we have
\eq{
D_{n'}|n;\bar k_1\eta_1,\ldots,\bar k_{\bar K}\eta_{\bar K}\rangle_a=0,
\qquad
1\le n\le\bar l,
\qquad
n'\ge1,
\label{ntphysbar}
}
subject to the equations
\eq{
\sum^{\bar K}_{m=1}{[\bar k_mn]\over[n]}\,\eta_m^n=0,
\qquad
1\le n\le\bar l,
\label{etaeqs}
}
are satisfied. We conclude that these vectors produce the Weyl invariant matrix elements,
\Multline{
{}_a\langle n;k_1\xi_1,\ldots,k_K\xi_K|t_{\kappa_1}(x_1)\ldots t_{\kappa_M}(x_M)
|n';\bar k_1\eta_1,\ldots,\bar k_{\bar K}\eta_{\bar K}\rangle_a
\\
={}_{wa}\langle n;k_1\xi_1,\ldots,k_K\xi_K|t_{\kappa_1}(x_1)\ldots t_{\kappa_M}(x_M)
|n';\bar k_1\eta_1,\ldots,\bar k_{\bar K}\eta_{\bar K}\rangle_{wa},
\qquad
w\in\cW,
\label{ffWinv}
}
which are form factors of some descendant operators for $1\le n\le l$, $1\le n'\le\bar n$.

\begin{theorem}
For generic values of the parameter $a$ the vectors ${}_a\langle n;k_1\xi_1,\ldots,k_K\xi_K|$ with the condition~(\ref{xieqs}) span the whole space $\cDRphys_n$ for $0\le n\le l$, while the vectors $|n;\bar k_1\eta_1,\ldots,\bar k_{\bar K}\eta_{\bar K}\rangle_a$ with~(\ref{etaeqs}) span the space $\cDLphys_n$ for $0\le n\le\bar l$.
\end{theorem}

Indeed, consider the bra-vectors. Subject to the condition (\ref{xieqs}), the coefficients $\kappa^{(s)}_n$ satisfy the equation
$$
\sum^L_{s=1}\kappa^{(s)}_n=0.
$$
It means that the r.~h.~s.\ of Eq.~(\ref{nXiform}) only contains differences of $d^{(i+1)}_n-d^{(i)}_n$ as it must be. Besides, this r.~h.~s.\ only depends on $L-1$ parameters for a given~$n$, e.~g.\ $\kappa^{(i)}_n$, $i=1,\ldots,L-1$. Now the same argumentation persuades us that these remaining $\kappa$s may be considered as independent variables, and the same argumentation proves that there is no vector orthogonal to the set of monomials. Then the same deformation argument should be applied. We arrive to the

\begin{theorem}
For any $l$ there exists an analytic in $a$ family of sets $\{h^{\text{\rm inv}}_{a,l,\mu}\in\cA_l\}^{\dim\cA_l}_{\mu=1}$, which are bases in $\cA_l$ for generic values of $a$, such that $r_a(w)h^{\text{\rm inv}}_{a,l,\mu}=h^{\text{\rm inv}}_{wa,l,\mu}$.
\end{theorem}

This alternative derivation has two advantages. First, it proves the existence of an \emph{analytic} in $a$ Weyl invariant basis in the space of the operators~$V^g_a$. Second, it provides a prescription to get form factors of these basic elements \emph{independently of the representation} for form factors in use. Moreover, it is easy to see, that it is not necessary to solve the equations (\ref{xieqs}) and (\ref{etaeqs}) explicitly. As it is shown in Appendix~\ref{app:EqSolutions} the form factors are rationally expressed in terms of an appropriately chosen set of independent variables. The Appendix provides a constructive way to obtain the form factors in terms of the independent variables.

\section{The simplest example: level $(1,0)$ descendants}
\label{sec:LevelOne}

Here we study the form factors of the level~$(1,0)$ operators. First, we find the recursion relations for these form factors and, second, we find the Weyl invariant combinations by means of the first approach from the previous section. Note that finding recursion relations is not necessary for obtaining the form factors explicitly, but recursion relations often turn out to be a useful tool to prove theorems.

As we have already mentioned, the whole reasoning (\ref{ResJ})--(\ref{J0infdif}) is valid for the $J$ functions related to arbitrary operators. Consider the function
\eq{
J^{\alpha_ic_{-1}}_{k,N+1,a}(z;X)
=\prod^N_{n=1}\prod_{m=1}^{k-1}f^{-1}\left({z\over x_n}\omega^{k+1-2m\over2}\right)\>
{}_a\langle1|\pi_R(\alpha_ic_{-1})t_k(z)t_1(x_1)\ldots t_1(x_N)|1\rangle_a.
\label{J(1,0)desc}
}
It admits the expansions (\ref{Jinfexp}) and (\ref{J0exp}) with appropriate functions $J^{(0)}$ and~$J^{(\infty)}$. Therefore, to obtain the recurrent relations for this function it is sufficient to find the asymptotics as $z\to0$ or $z\to\infty$.

Rewrite the expression (\ref{J(1,0)desc}) in the form
\Multline{
J^{\alpha_ic_{-1}}_{k,N+1,a}(z,X)
\prod^N_{n=1}\prod_{m=1}^{k-1}f\left({z\over x_n}\omega^{k+1-2m\over2}\right)
={}_a\langle1|[\pi_R(\alpha_ic_{-1}),t_k(z)]t_1(x_1)\ldots t_1(x_N)|1\rangle_a
\\*
+{}_a\langle1|t_k(z)\pi_R(\alpha_ic_{-1})t_1(x_1)\ldots t_1(x_N)|1\rangle_a.
\label{J(1,0)commute}
}
From (\ref{piRcommut}) we have
$$
[\pi_R(\alpha_ic_{-1}),t_k(z)]
=z\sum_{1\leq s_1<\ldots<s_k\leq L}\omega^{\langle a,H_{s_1,\ldots,s_k}\rangle}
\sum_{n=1}^k\langle\alpha_i,H_{s_n}\rangle\,\omega^{-{L-k-2(s_n-n)\over 2}}
\prod_{m=1}^k\lambda_{s_m}(z\omega^{k+1-2m\over 2}).
$$
Since $f(z)=1+O(z^{-1})$ as $z\to\infty$, the linear terms in the $z$ expansion of the function $J^{\alpha_jc_{-1}}_{k,N+1,a}(z,X)$ only come from the first term in the r.~h.~s.\ of Eq.~(\ref{J(1,0)commute}):
$$
J^{\alpha_ic_{-1}}_{k,N+1,a}(z,X)
=zK^i_{k,a}J_{1,N,a}(X)+O(z^0)
$$
with
\eq{
K^i_{k,a}
=\sum_{1\leq s_1<\ldots<s_k\leq L}\omega^{\langle a,H_{s_1,\ldots,s_k}\rangle}
\sum_{n=1}^k\langle\alpha_i,H_{s_n}\rangle\,\omega^{-{L-k-2(s_n-n)\over 2}}.
\label{K1def}
}
It is not easy to find the term of the order $z^0$ as $z\to\infty$, but it is not necessary, since it is the leading term in the limit $z\to0$. From the cluster property we readily get
$$
J_{k,N+1,a}^{\alpha_ic_{-1}}(0,X)=K_{k,a}J_{1,N,a}^{\alpha_ic_{-1}}(X).
$$
Finally, we obtain the recurrent relation
\eq{
\Aligned{
J^{\alpha_ic_{-1}}_{k,N+1,a}(z;X)
&=K_{k,a}J_{1,N,a}^{\alpha_ic_{-1}}(X)+zK^i_{k,a}J_{1,N,a}(X)
\\*
&\quad-\sum^N_{n=1}{x^{-1}_n\omega^{-{k+1\over2}}\over z^{-1}-x_n^{-1}\omega^{-{k+1\over2}}}
\>R^+_{N,n}(X)J^{\alpha_ic_{-1}}_{k+1,N,a}(x_n\omega^{k\over2};\hat X_n)
\\*
&\quad+\sum^N_{n=1}{x^{-1}_n\omega^{{k+1\over2}}\over z^{-1}-x_n^{-1}\omega^{{k+1\over2}}}
\>R^-_{N,n}(X)J^{\alpha_ic_{-1}}_{k+1,N,a}(x_n\omega^{-{k\over2}};\hat X_n).
}\label{Jlevel1rec}
}
Together with the initial and cyclic conditions
\eq{
J^{\alpha_ic_{-1}}_{k,0,a}=0,
\qquad
J^{\alpha_ic_{-1}}_{L,N+1,a}(z;X)=J^{\alpha_ic_{-1}}_{1,N,a}(X)
\label{Jlevel1init}
}
and with the known recursion relation for $J_{k,N,a}(z;X)$ it uniquely defines the form factors of the level~$(1,0)$ descendant operators.

The next step is to find the combinations invariant with respect to the Weyl algebra. Evidently, it is sufficient to find the Weyl invariant $k$-independent combinations of the functions~$K^i_{k,a}$. But it is very difficult to find these combinations by trial and error. We shall better find them using the technique described in the first part of Sec.~\ref{sec:ReflectionDescendant}. Namely, consider the expansion of the product ${}_a\langle1|t_k(z)$ up to the power $z^{-1}$ as $z\to\infty$. We have
$$
{}_a\langle1|t_k(z)={}_a\langle1|K_{k,a}+z^{-1}{}_a\langle1;k|+O(z^{-2})
$$
with
\eq{
{}_a\langle1;k|
=\sum_{1\le s_1<\cdots<s_k\le L}\omega^{\langle a,H_{s_1\ldots s_k}\rangle}
\sum^k_{m=1}\omega^{m-{k+1\over2}}\>{}_a\langle1|d_1^{(s_m)}.
\label{1kstate}
}
It is straightforward to check that
$$
{}_a\langle1;k|D_{-1}
={}_a\langle1|\omega^{1/2}A^+_1[k]K_{k,a}.
$$
Introduce the state
\eq{
{}_a\langle C|
={1\over\omega^{L+1\over2}A^+_1}\left(
\sum^{L-1}_{k=1}{c_k\over[k]K_{k,a}}\>{}_a\langle1;k|
+c_L\>{}_a\langle1;L|
\right),
\qquad
C=(c_1,\ldots,c_L).
\label{Cstate}
}
This state is `physical' if ${}_a\langle C|D_{-1}=0$, which takes place if
\eq{
\sum^{L-1}_{k=1}c_k=0.
\label{sumc0}
}
This equation admits $L-1$ independent solution $C^{(\sigma)}$, $\sigma=1,\ldots,L-1$. To identify these solutions to some elements of $\cA$ we have to write down the elements $d^{(s)}_1$, $s>1$ in the form
$$
d^{(s)}_1=d^{(1)}_1-\sum^{s-1}_{i=1}(d^{(i)}_1-d^{(i+1)}_1)
=d^{(1)}_1-A^+_1\sum^{s-1}_{i=1}\omega^{L+1-2i\over2}R^{(i)}_1.
$$
After substituting it into (\ref{1kstate}), (\ref{Cstate}) the coefficient at the element $d^{(1)}_1$ cancels out due to~(\ref{sumc0}).

Let us single out one solution:
\eq{
c^{(L-1)}_1=\cdots=c^{(L-1)}_{L-1}=0,
\qquad
c^{(L-1)}_L=1.
\label{C(L-1)solution}
}
This solution corresponds to the integral of motion
\eq{
{}_a\langle C^{(L-1)}|={}_a\langle\iota_1|.
\label{C(L-1)=iota1}
}
Now, let $C^{(\sigma)}$, $\sigma=1,\ldots,L-2$ be any basis in the subspace of the solutions to the equation (\ref{sumc0}) with $c_L=0$. Then ${}_a\langle C^{(\sigma)}|={}_a\langle h_{1,a}^{(\sigma)}|$, where
\eq{
h_{1,a}^{(\sigma)}
=-\sum^{L-1}_{i=1}\alpha_ic_{-1}
\sum^{L-1}_{k=1}{c^{(\sigma)}_k\over[k]K_{k,a}}
\sum^k_{m=1}\omega^{m-i-{k+1\over2}}
\sum_{1\le s_1<\cdots<s_k\le L\atop s_m>i}\omega^{\langle a,H_{s_1\ldots s_k}\rangle},
\qquad
\sigma=1,\ldots,L-2.
\label{level1inv}
}
In particular, $h^{(L-1)}_{1,a}=\iota_1$ assuming~(\ref{C(L-1)solution}).

If we only allow $a$-independent solutions $C^{(\sigma)}$, the coefficients in the linear combination (\ref{Cstate}) are Weyl invariant. Hence, by the construction
$$
{}_a\langle h_{1,a}^{(\sigma)}|t_{k_1}(x_1)\ldots t_{k_N}(x_N)|1\rangle_a
={}_{wa}\langle h_{1,wa}^{(\sigma)}|t_{k_1}(x_1)\ldots t_{k_N}(x_N)|1\rangle_{wa}
\quad\forall w\in\cW
$$
for $\sigma=1,\ldots,L-1$.

\section{Discussion}

In this paper we extended the main results of~\cite{Feigin:2008hs} to the case of the $A^{(1)}_{L-1}$ affine Toda models. We construct spaces of solutions to the form factor bootstrap equations, which, as we argue, can be bijectively mapped onto the Fock spaces of descendant operators over the exponential operators~$V_a(x)$ for generic values of~$a$. We propose a construction to find Weyl invariant families of bases in these spaces based on high rapidity asymptotic expansions of the form factors of exponential operators. In principle, it is possible, at least for the lowest levels, to obtain Weyl invariant families of bases in the Fock spaces of descendant operators in the Lagrangian formalism\cite{Fateev:1998xb}. However, the identification of both types of bases cannot be unique without some additional information. Probably, we could fix identification at some special resonant points, but it has not been done up to now. Thus, the field identification problem of the bootstrap form factor program remains unsolved.

Recently, in the remarkable papers~\cite{Boos:2009fs,Jimbo:2009ja} it was shown using the scaling limit from a lattice model, that the spaces of descendant operators, at least in the case of the sine-Gordon model, can be created by use of some fermionic operators acting in the space of local operators of the theory. In particular, it turned out to be possible to calculate exactly all the expectation values of descendant operators in the theory\cite{Jimbo:2009ja}. It would be utterly unnatural, if such fermionic operators would not induce an action on the algebra $\cA\otimes\cA$ in our construction. Hence, revealing such fermions in a construction for form factors would be an important step toward the field identification, if not its complete solution.

\section*{Acknowledgments}

The authors are grateful to M.~Bershtein, B.~Feigin, Ya.~Pugai and F.~Smirnov for stimulating discussions. The work was supported, in part, by RFBR under the grants 08--01--00720 and 09--02--12446, by RFBR and CNRS under the grant 09--02--93106, by the Program for Support of Leading Scientific Schools under the grant 3472.2008.2 and by the Federal Program ``Scientific and Scientific-Pedagogical Personnel of Innovational Russia'' under the state contract No.~P1339. The visit of M.~L.\ to LPTHE, Universit\'e Paris~6 in August--September of 2009, where a part of the work was done, was supported by CNRS in the framework of the LIA ``Physique Th\'eorique et Mati\`ere Condens\'ee'' (ENS--Landau program).

\Appendix

\section{Simple poles of $t_k(z)t_1(x)$}
\label{app:simplepoles}

Here we prove that the poles (\ref{tkt1SimplePoles}) are simple. The product $t_k(z)t_1(x)$ could possess double poles at the points $z=x\omega^{\pm(k-1)/2}$ due to two poles of the function~$f(z)$. For example, for the pole $z=x\omega^{-(k-1)/2}$ the double poles appear in the two types of terms. Let us obtain the terms of the first type. Let $\sigma_1$, $\sigma_2$ be two integers such that $2\le\sigma_1<\sigma_2\le L$. Then the expression (\ref{tndef}) for $t_k$ contains terms with $s_{k-1}=\sigma_1-1$ and $s_k=\sigma_1$, while that for $t_1$ contains a term with $s=\sigma_2$. Hence, the product $t_k(z)t_1(x)$ contains terms of the form
$$
\lcolon\lambda_*\lambda_{\sigma_1-1}\left(z\omega^{3-k\over2}\right)
\lambda_{\sigma_1}\left(z\omega^{1-k\over2}\right)\lambda_{\sigma_2}(x)\rcolon\>
f\left({z\over x}\omega^{3-k\over2}\right)f\left({z\over x}\omega^{1-k\over2}\right),
$$
where $\lambda_*$ means the product of all other $\lambda$s and~$f$s. The product of the two $f$ functions produces a double pole at the point $z=x\omega^{-{k+1-2i\over2}}$. In the vicinity of this point it behaves as
\eq{
\lcolon\lambda_*\lambda_{\sigma_1-1}(\omega x)\lambda_{\sigma_1}(x)\lambda_{\sigma_2}(x)\rcolon\>
{-(1-\omega^{1-p})^2(1-\omega^p)^2\over(1-\omega)^2\left({z\over x}\omega^{1-k\over2}-1\right)^2}
+O\left(\left({z\over x}\omega^{1-k\over2}-1\right)^{-1}\right).
\label{doublepoleterm}
}
The second type of terms consists of those with $s_{i-1}=\sigma_1-1$, $s_i=\sigma_2$, $s=\sigma_1$:
$$
\lcolon\lambda_*\lambda_{\sigma_1-1}\left(z\omega^{3-k\over2}\right)
\lambda_{\sigma_2}\left(z\omega^{1-k\over2}\right)\lambda_{\sigma_1}(x)\rcolon\>
f\left({z\over x}\omega^{3-k\over2}\right)f\left({x\over z}\omega^{-{1-k\over2}}\right).
$$
It possesses a double pole at the same point, where it behaves just as minus the expression~(\ref{doublepoleterm}). It is easy to see that the factors denoted by $\lambda_*$ are the same for both expressions in this limit if all other $s_j$ coincide. Hence, both double pole contributions cancel each other. The same reasoning is valid for the pole at $z=x\omega^{(k-1)/2}$.

\section{Equation of motion}
\label{app:EquationOfMotion}

In this appendix we prove that form factors are consistent with the equation of motion
$$
\alpha_i\d\bd\varphi
={\pi\mu b\over2}(2e^{b\alpha_i\varphi}-e^{b\alpha_{i-1}\varphi}-e^{b\alpha_{i+1}\varphi}).
$$
The derivatives of a field produce multiplication of its form factors by the components of the momentum according to the usual rule $P_\mu\leftrightarrow i\d_\mu$. Introduce the notation
$$
S^k_n(z;X)={\sin{\pi kn\over L}\over\sin{\pi n\over L}}z^n+S_n(X).
$$
Let $z=\e^\theta$, $x_n=\e^{\theta_n}$. Then the components of the momentum are given by
$$
\Aligned{
P_z(\theta,\theta_1,\ldots,\theta_N)_{k,1,\ldots,1}
=-{m\over2}S_1^k(z;X)
\\
P_\bz(\theta,\theta_1,\ldots,\theta_N)_{k,1,\ldots,1}
={m\over2}S_{-1}^k(z;X).
}
$$
Let
$$
a=\sum^{L-1}_{i=1}a_i\alpha_i,
\qquad
\nu_i=p\alpha_i-\rho.
$$
Then we have
$$
\lvac\alpha_i\d\bd\varphi|\theta,\theta_1,\ldots,\theta_N\rangle_{k,1,\ldots,1}
={m^2\over4Q}S^k_1(z;X)S^k_{-1}(z;X)
\left.{d\over d a_i}f_a(\theta,\theta_1,\ldots,\theta_N)_{k1\ldots1}\right|_{a=-\rho}
$$
and
$$
\lvac\e^{b\alpha_i\varphi}|\theta,\theta_1,\ldots,\theta_N\rangle_{k1\ldots1}
=\omega^{\mp(k+N)}\lvac\e^{b\alpha_{i\pm1}\varphi}|\theta,\theta_1,\ldots,\theta_N\rangle_{k1\ldots1}
=G_{\nu_i}f_{\nu_i}(\theta,\theta_1,\ldots,\theta_N)_{k1\ldots1}.
$$
All values $G_{\nu_i}$ are evidently equal. Let
$$
J'_{k,N+1,i}(z;X)=\left.{d\over d a_i}J_{k,N+1,a}(z;X)\right|_{a=-\rho}\>.
$$
The equation of motion can be rewritten as
\eq{
S_1^k(z;X)S_{-1}^k(z;X)J'_{k,N+1,i}(z;X)=A\sin^2{\pi(k+N)\over L}J_{k,N+1,\nu_i}(z;X),
\label{EqMotion}
}
where
$$
A={8\pi\mu G_{\nu_i}\over(1-p)m^2}.
$$
According to \cite{Fateev:2001mj} it reads
\eq{
A={\pi\over L\sin{\pi\over L}\sin{\pi p\over L}\sin{\pi(1-p)\over L}}.
\label{Adef}
}
It is easy to check Eq.~(\ref{EqMotion}) with (\ref{Adef}) for $N=0$ by using (\ref{Jinit}),~(\ref{Kspec}). Now we prove it for $N>0$ by induction.

It is more convenient to use induction in the variable $k+N$ rather than just~$N$. Suppose that the equation (\ref{EqMotion}) is valid for some value $M=k+N$ for arbitrary $k=1,\ldots,L-1$. Taking derivatives of both sides of the recurrent relation (\ref{Jrec}) we get
$$
\Aligned{
J'_{k,N+1,i}(z;X)=
&\quad\sum^N_{n=1}{x_n\omega^{k+1\over2}\over z-x_n\omega^{k+1\over2}}
\>R^+_{N,n}(X)J'_{k+1,N,i}(x_n\omega^{k\over2};\hat X_n)
\\*
&\quad-\sum^N_{j=1}{x_n\omega^{-{k+1\over2}}\over z-x_n\omega^{-{k+1\over2}}}
\>R^-_{N,n}(X)J'_{k+1,N,i}(x_n\omega^{-{k\over2}};\hat X_n).
}
$$
Multiplying it by $S_1^k(z,X)S_{-1}^k(z,X)$ and using the identity~(\ref{RRidentity}) we get
\Align{
&S_1^k(z,X)S_{-1}^k(z,X)J'_{k,N+1}(z;X)
=\sum^N_{n=1}[k] x_n\omega^{k+1\over2}\>R^+_{N,n}(X)
S_{-1}^{k+1}(x_n\omega^{k\over2};\hat X_n)J'_{k+1,N}(x_n\omega^{k\over2};\hat X_n)
\notag
\\*
&\qquad +\sum^N_{n=1}{x_n\omega^{k+1\over2}\over z-x_n\omega^{k+1\over2}}\>R^+_{N,n}(X)
S_1^{k+1}(x_n\omega^{k\over2};\hat X_n)S_{-1}^{k+1}(x_n\omega^{k\over2};\hat X_n)
J'_{k+1,N}(x_n\omega^{k\over2};\hat X_n)
\notag
\\*
&\qquad-\sum^N_{n=1}[k] x_n\omega^{-{k+1\over2}}\>R^-_{N,n}(X)
S_{-1}^{k+1}(x_n\omega^{-{k\over2}};\hat X_n)J'_{k+1,N}(x_n\omega^{-{k\over2}};\hat X_n)
\notag
\\*
&\qquad-\sum^N_{n=1}{x_n\omega^{-{k+1\over2}}\over z-x_n\omega^{-{k+1\over2}}}\>R^-_{N,n}(X)
S_1^{k+1}(x_n\omega^{-{k\over2}};\hat X_n)S_{-1}^{k+1}(x_n\omega^{-{k\over2}};\hat X_n)
J'_{k+1,N}(x_n\omega^{-{k\over2}};\hat X_n).
\label{IndRecRel}
}
Due to the induction hypothesis we have
$$
S_1^{k+1}(x_n\omega^{-{k\over2}};\hat X_n)S_{-1}^{k+1}(x_n\omega^{-{k\over2}};\hat X_n)
J'_{k+1,N,i}(x_n\omega^{-{k\over2}};\hat X_n)
=A\sin^2{\pi(k+N)\over L}J_{k+1,N,\nu_i}(x_n\omega^{-{k\over2}};\hat X_n).
$$
Hence, the sum of the first and the third terms in the right hand side of Eq.~(\ref{IndRecRel}) are equal to
\Align{
&A\sin^2{\pi(k+N)\over L}\Biggl(
\sum^N_{n=1}{[k] x_n\omega^{k+1\over2}\over S_1^{k+1}(x_n\omega^{k\over2};\hat X_n)}\>
R^+_{N,n}(X)J_{k+1,N,\nu_i}(x_n\omega^{k\over2};\hat X_n)
\notag
\\
&\qquad\qquad\qquad
-\sum^N_{n=1}{[k] x_n\omega^{-{k+1\over2}}\over S_1^{k+1}(x_n\omega^{-{k\over2}};\hat X_n)}\>
R^-_{N,n}(X)J_{k+1,N,\nu_i}(x_n\omega^{-{k\over2}};\hat X_n)
\Biggr)
\notag
\\
&\qquad
=-A\sin^2{\pi(k+N)\over L}\>
(J_{k,N+1,\nu_i}(-[k]^{-1}S_1(X);X)-K_{k,\nu_i}J_{1,N,\nu_i}(X)),
\notag
}
while the two remaining terms reads
\Align{
&A\sin^2{\pi(k+N)\over L}\Biggl(
\sum^N_{n=1}{x_n\omega^{k+1\over2}\over z-x_n\omega^{k+1\over2}}
\>R^+_{N,n}(X)J_{k+1,N,\nu_i}(x_n\omega^{k\over2};\hat X_n)
\notag
\\
&\qquad\qquad\qquad
-\sum^N_{n=1}{x_n\omega^{-{k+1\over2}}\over z-x_n\omega^{-{k+1\over2}}}
\>R^-_{N,n}(X)J_{k+1,N,\nu_i}(x_n\omega^{-{k\over2}};\hat X_n)
\Biggr)
\notag
\\
&\qquad
=A\sin^2{\pi(k+N)\over L}\>
(J_{k,N+1,\nu_i}(z;X)-K_{k,\nu_i}J_{1,N,\nu_i}(X)).
\notag
}
Gathering these terms we get
$$
S_1^k(z;X)S_{-1}^k(z;X)J'_{k,N+1,i}(z;X)
=A\sin^2{\pi(k+N)\over L}\>
(J_{k,N+1,\nu_i}(z;X)-J_{k,N+1,\nu_i}(-[k]^{-1}S_1(X);X)).
$$
It is nearly what we need. To prove the equation of motion for $k+N=M+1$ it remains to prove that
\eq{
J_{k,N+1,\nu_i}(-[k]^{-1}S_1(X);X)=0.
\label{Jextra}
}
Consider first the case $k=1$. Then the function $J_{1,N+1,\nu_i}(x_{N+1};x_1,\ldots,x_N)$ is symmetric with respect to all of the variables $x_1,\ldots,x_{N+1}$. Hence, any of these variables can be chosen for~$z$. It means that
$$
J_{1,N+1,\nu_i}(-S_1(X);X)=J_{1,N+1,\nu_i}(-S_1(x_{N+1},\hat X_j);x_{N+1},\hat X_j).
$$
The left hand side is $x_N$-independent, while the right hand side is $x_j$-independent. It means that the function $J_{1,N+1,\nu_i}(-S_1(X);X)$ is constant in its all $N$ variables. Therefore, it is sufficient to prove that it is zero e.~g.\ for $x_N\to\infty$. Let us use the recurrent relation
$$
\Aligned{
J_{1,N+1,\nu_i}(-S_1(X);X)
&=K_{1,\nu_i}J_{1,N,\nu_i}(X)
\\
&\quad+\sum^N_{n=1}{x_n\omega\over S_1^2(x_n\omega^{1\over2};\hat X_n)}\>
R^+_{N,n}(X)J_{2,N,\nu_i}(x_n\omega^{1\over2};\hat X_n)
\\
&\quad-\sum^N_{n=1}{x_n\omega^{-1}\over S_1^2(x_n\omega^{-{1\over2}};\hat X_n)}\>
R^-_{N,j}(X)J_{2,N,\nu_i}(x_n\omega^{-{1\over2}};\hat X_n).
}
$$
Since the left hand side is a constant, we may calculate it in the limit $x_N\to\infty$. In this limit the only nonvanishing terms in the sums in the right hand side are those with $n=N$. Taking into account that
$$
\>R^+_{N,n}(X)=\>R^-_{N,n}(X)
=-{2\i\sin{\pi p\over L}\sin{\pi(1-p)\over L}\over \sin{\pi\over L}}+O(x_N^{-1})
\text{ as $x_N\to\infty$},
$$
we obtain
$$
J_{1,N+1,\nu_i}(-S_1(X);X)
\to\left((K_{1,\nu_i})^2
+4{\sin{\pi\over L}\sin{\pi p\over L}\sin{\pi(p-1)\over L}\over\sin{2\pi\over L}}K_{2,\nu_i}
\right)J_{1,N-1,\nu_i}(\hat X_N)=0.
$$
Hence, $J_{1,M,\nu_i}(-S_1(X);X)=0$. Now, it is straightforward to check (\ref{Jextra}) by fusing $t_1$'s into $t_k$. That is why we used induction in the variable $k+N$ rather than~$N$.

\section{Solutions to Eqs.~(\ref{xieqs}), (\ref{etaeqs}) and form factors}
\label{app:EqSolutions}

As the equations (\ref{xieqs}) and (\ref{etaeqs}) have the same form, we restrict the consideration to the bra-vectors.

Since $t_k(z)$ for $k=2,\ldots,L-1$ can be obtained by the fusion of $t_1(z)$ according to~(\ref{Restkt1}), it is sufficient to consider $t_1(z)$ and $t_L(z)$ without loss of generality. Besides, since the differences $t_L(z\omega)-t_L(z)$ also appear in such fusion according to~(\ref{Restkt1kin}), it is convenient to consider the `symmetrized' version of $t_L(z)$:
\eq{
\tLsym(z)=\sum^{L-1}_{m=0}t_L(z\omega^m),
\label{tLsymdef}
}
which is an operator valued function of~$z^L$. Similarly to (\ref{ntstates}) consider the expansion
\eq{
{}_a\langle1|t_1(\xi_1^{-1}z)\ldots t_1(\xi_r^{-1}z)
\tLsym(\zeta_1^{-1/L}z)\ldots\tLsym(\zeta_s^{-1/L}z)
=\sum^\infty_{n=0}z^{-n}\>{}_a\langle n;\xi_1,\ldots,\xi_r;\zeta_1,\ldots,\zeta_s|.
\label{brasymdef}
}
The equations (\ref{xieqs}) reduce in this case to
\eq{
\Aligned{
\sum^r_{m=1}\xi_m^n
&=0,
\qquad
n\not\in L\Z,
\\
\sum^r_{m=1}\xi_m^n
&=-L\sum^s_{m=1}\zeta_m^{n/L}=\Sigma_{n/L},
\qquad
n\in L\Z,
}\label{xieqssym}
}
for $1\le n\le l$. Here we introduced the variables $\Sigma_\nu$, $\nu=1,\ldots,\lambda=\lfloor l/L\rfloor$, which will be useful below. Due to the Newton--Girard identities, for $1\le n\le l$ the quantities $\sigma^1_n$ vanish, if $n\not\in L\Z$, while $\sigma^1_L,\sigma^1_{2L},\ldots,\sigma^1_{\lambda L}$ are in one-to-one correspondence with $\Sigma_1,\ldots,\Sigma_\lambda$. The quantities $\sigma^L_n$ for $n=1,\ldots,\lambda$ are also in one-to-one correspondence with $\Sigma_1,\ldots,\Sigma_\lambda$ and, hence, with $\sigma^1_L,\sigma^1_{2L},\ldots,\sigma^1_{\lambda L}$.

The form factors
$$
{}_a\langle n;\xi_1,\ldots,\xi_r;\zeta_1,\ldots,\zeta_s|t_{k_1}(x_1)\ldots t_{k_N}(x_N)|h'\rangle
$$
are rational symmetric functions of the variables $\xi_1,\ldots,\xi_r$ and the variables $\zeta_1,\ldots,\zeta_s$, that is they are ratios of symmetric polynomials. Therefore, if we learn how to calculate elementary symmetric polynomials $\sigma^1_n=\sigma_n(\xi_1,\ldots,\xi_r)$, $n=1,\ldots,r$, and $\sigma^L_n=\sigma_n(\zeta_1,\ldots,\zeta_s)$, $n=1,\ldots,\lambda$ on solution of the equations (\ref{xieqssym}), we will be able to calculate the form factors.

We have $r+s$ variables and $l$ equations, that is $r+s-l$ independent variables. Let $r_0=r-l+\lambda$ and $s_0=s-\lambda$ so that $r_0+s_0=r+s-l$. Take $\xi_1,\ldots,\xi_{r_0}$ and $\zeta_1,\ldots,\zeta_{s_0}$ for independent variables. Then the variables $\xi_1,\ldots,\xi_r$ and $\zeta_1,\ldots,\zeta_s$ are solutions to the equations
\Align{
\xi_m^r+\sum^\lambda_{n=1}(-)^{Ln}\sigma^1_{Ln}\xi_m^{r-Ln}
+\sum^r_{n=l+1}(-)^n\sigma^1_n\xi_m^{r-n}
&=0,
\label{sigma1eqs}
\\
\zeta_m^s+\sum^s_{n=1}(-)^n\sigma^L_n\zeta_m^{r-n}
&=0.
\label{sigmaLeqs}
}
Consider the equations~(\ref{sigma1eqs}) for $m=1,\ldots,r_0$ as a system of $r_0$ linear inhomogeneous equations for $r_0$ variables $\sigma^1_L,\sigma^1_{2L},\ldots,\sigma^1_{\lambda L},\sigma^1_{l+1},\sigma^1_{l+1},\ldots,\sigma^1_r$. For generic values of $\xi_1,\ldots,\xi_{r_0}$ these equations are nondegenerate and are solved in terms of the Schur polynomials. Now, using the Newton--Girard identities we can express $\sigma^L_1,\ldots,\sigma^L_\lambda$ as polynomials of $\sigma^1_L,\ldots,\sigma^1_{\lambda L}$. Thus, the equations (\ref{sigmaLeqs}) for $m=1,\ldots,s_0$ become $s_0$ linear inhomogeneous equations for $s_0$ variables $\sigma^L_{\lambda+1},\ldots,\sigma^L_s$, which can be solved in terms of the Schur polynomials as well.

Finally, we expressed all the symmetric polynomials $\sigma^1_n$, $\sigma^L_n$ and, hence, the form factors as rational functions of the independent variables $\xi_1,\ldots,\xi_{r_0},\zeta_1,\ldots,\zeta_{s_0}$, q.~e.~d. Note that though for simplicity we omitted some explicit formulas, the procedure described here is thoroughly constructive.

\raggedright

\end{document}